\documentclass[fleqn,usenatbib]{mnras}

\usepackage[T1]{fontenc}
\usepackage{ae,aecompl}

\DeclareRobustCommand{\VAN}[3]{#2}
\let\VANthebibliography\thebibliography
\def\thebibliography{\DeclareRobustCommand{\VAN}[3]{##3}\VANthebibliography}

\usepackage{graphicx}	
\usepackage{amsmath}	
\usepackage{amssymb}	
\usepackage{pdflscape}
\usepackage{pdfpages}
\usepackage{adjustbox}
\usepackage{multirow}
\usepackage[usestackEOL]{stackengine}
\usepackage{ulem}

\usepackage{newtxtext,newtxmath}

\newcommand{\kms}{\,km\,s$^{-1}$}
\newcommand{\ms}{M$_{\sun}$}
\definecolor{m-purple}{HTML}{7e1e9c}
\definecolor{m-green}{rgb}{0.0, 0.5, 0.0}

\newcommand{\CFe}{\mbox{$\mbox{[C/Fe]}$}}
\newcommand{\aFe}{\mbox{$\mbox{[$\alpha$/Fe]}$}}
\newcommand{\Mh}{\mbox{$\mbox{[M/H]}$}}
\newcommand{\Zh}{\mbox{$\mbox{[Z/H]}$}}
\newcommand{\Feh}{\mbox{$\mbox{[Fe/H]}$}}

\newcommand{\aM}{\mbox{$\mbox{[$\alpha$/M]}$}}
\newcommand{\cM}{\mbox{$\mbox{[C/M]}$}}
\newcommand{\liH}{\mbox{$\mbox{[Li/H]}$}}
\newcommand{\OFe}{\mbox{$\mbox{[O/Fe]}$}}

\newcommand{\logg}{$\log g$}
\newcommand{\Teff}{${\rm T}_{\mbox{\scriptsize eff}}$}

\title[CO Absorption Features in ETGs]{Strong CO Absorption Features in Massive ETGs}

\author[E. Eftekhari et al.]{
Elham Eftekhari,$^{1,2}$\thanks{E-mail: elhamea@iac.es }
Francesco La Barbera,$^{3}$
Alexandre Vazdekis,$^{1,2}$
Carlos Allende Prieto,$^{1,2}$ 
\newauthor
Adam Thomas Knowles$^{4}$
\\
$^{1}$Instituto de Astrof{\'{\i}}sica de Canarias, E-38200 La Laguna, Tenerife, Spain\\
$^{2}$Departamento de Astrof{\'{\i}}sica, Universidad de La Laguna, E-38205 La Laguna, Tenerife, Spain\\
$^{3}$INAF-Osservatorio Astronomico di Capodimonte, sal. Moiariello 16, Napoli I-80131, Italy\\
$^{4}$Jeremiah Horrocks Institute, School of Physical Sciences and Computing, University of Central Lancashire, Preston, PR1 2HE, UK
}

\date{Accepted 2022 February 17. Received 2022 February 17; in original form 2021 July 14}

\pubyear{2021}

\begin{document}
\label{firstpage}
\pagerange{\pageref{firstpage}--\pageref{lastpage}}
\maketitle

\begin{abstract}

Massive Early-Type Galaxies (ETG) in the local Universe are believed to be the most mature stage of galaxy evolution. Their stellar population content reveals the evolutionary history of these galaxies. However, while state-of-the-art Stellar Population Synthesis (SPS) models provide an accurate description of observed galaxy spectra in the optical range,  the modelling in the Near-Infrared (NIR) is still in its infancy. Here we focus on NIR CO absorption features to show, in a systematic and comprehensive manner, that for massive ETGs, all CO indices, from \textit{H} through to \textit{K} band, are significantly stronger than currently predicted by SPS models. We explore and discuss several possible explanations of this ``CO mismatch'', including the effect of intermediate-age, AGB-dominated, stellar populations, high metallicity populations, non-solar abundance ratios and the initial mass function. While none of these effects is able to reconcile models and observations, we show that ad-hoc ``empirical'' corrections, taking into account the effect of CO-strong giant stars in the low-temperature regime, provide model predictions that are closer to the observations. Our analysis points to the effect of carbon abundance as the most likely explanation of NIR CO line-strengths, indicating possible routes for improving the SPS models in the NIR.

\end{abstract}




\section{Introduction} \label{sec:intro}



The study of the stellar content of galaxies, both in the local and early Universe, is fundamental to understand how they shape-up over cosmic time, as it provides key constraints on star formation rates, total stellar masses, chemical enrichment, and the stellar Initial Mass Function (IMF). Since Stellar Population Synthesis (SPS) models are an essential tool to constrain the stellar content of galaxies, a detailed assessment of their validity and limitations is crucial to determine the physical and evolutionary properties of these systems. While spectral synthesis modelling at optical wavelengths is now a mature field of research and optical galaxy spectra can be accurately matched with SPS models, there is still a long road ahead for Near-Infrared (NIR) SPS models to consistently agree with observations~\citep{riffel2019, eftekhari2021}. For example, only in the last decade, the problem of matching strong sodium absorption lines of massive Early-Type Galaxies (ETGs) has been scrutinized in the NIR. The strength of NIR sodium features in massive ETGs is much stronger than would be expected from an old stellar population with a Milky Way (MW)-like IMF and with solar elemental abundance ratios. A combination of a bottom-heavy IMF and a highly-enhanced sodium abundance can reconcile the tension between observations and NIR models \citep{smith2015, labarbera2017, rock2017}; however, the finding of massive ETGs with MW-like IMFs -as derived by strong gravitational lensing analyses- and strong sodium line-strengths at 1.14~$\mu$m calls for caution in interpreting the NIR sodium line-strengths \citep{smith2013, smith2015}. Another disagreement between observations and stellar population models in the NIR arises from CO absorption features, that are prominent in the \textit{H}- and \textit{K}-band spectral regions, and have always been a puzzle. 

The appearance of the first overtone of CO in \textit{K} band, at $\sim2.3~\mu$m, in the spectra of galaxies was discussed by many authors in the 1970s and 1990s \citep{baldwin1973b, frogel1975, frogel1978, aaronson1978, frogel1980, oliva1995, mobasher1996, james1999}.  CO absorption originates in the atmospheres of red giants and supergiants, that tend to have deeper CO absorptions than dwarf stars \citep{baldwin1973a}.  \citet{faber1972} opened up the discussion that optical data could not be used to uniquely determine the proportion of M dwarfs and M giants in the galactic nucleus of M31, showing that while a model with enhanced M dwarfs in synthesised models would match the Na doublet at $8190$~\AA, a model dominated by M giants is required to explain the \textit{K}-band CO strength. Since then, several authors have analyzed the \textit{K}-band CO absorption of galaxies, alone or in combination with other indices, by comparing the observed strengths with those of stellar spectra~\citep{baldwin1973a, baldwin1973b, frogel1975, frogel1978, oliva1995, mobasher1996, james1999}. All of these studies found that line-strengths of the \textit{K}-band CO absorption lie in the range of giant stars, concluding that most of the light emitted from galaxies in the CO spectral region comes from these stars. 

Using NIR observations of globular clusters, \citet{aaronson1978} showed that the $2.3~\mu$m CO index strength is strongly correlated with metallicity. They constructed SPS models and compared them with observations of the central regions of ETG, claiming that metal-rich models with a Salpeter IMF adequately fit the CO index of the brightest ellipticals. \citet{frogel1978} also found that any significant increase in the number of late-type dwarfs beyond those already contained in the SPS models drives the \textit{K}-band CO index to unacceptably low values\footnote{\citet{kroupa1994} obtained a similar result by simulating the \textit{K}-band spectrum of the cooling-flow ellipticals, i.e elliptical galaxies with a low-mass star accretion population. They used the spectral library of low-mass stars from \citet{arnaud1989} and showed that by spectroscopy around the \textit{K}-band CO feature, an overabundance of low-mass stars in these galaxies can be estimated.}, concluding that the changes observed in the CO indices of galaxies are due to variations in the mean metallicity of their stellar populations. However, \citet{frogel1980} attributed  the differences between various colours and \textit{K}-band CO index of ETGs with respect to those of globular clusters and stellar synthesis models to a population of low-temperature luminous stars present neither in the clusters nor in the models. They hypothesized giant branch stars with higher metallicity than the Sun and/or asymptotic giant-branch (AGB) stars above the first red giant tip as two candidates for such a population. 

Separation of the relative contributions to the \textit{K}-band CO line-strength from young supergiants in a burst population (a few Myr) and giants in an older stellar system ($\sim1$~Gyr)  has also been a subject of debate; the NIR CO features are mainly sensitive to effective temperature but are also somewhat shallower in giants than supergiants of similar temperatures \citep{kleinmann1986, origlia1993, oliva1995}. However, metal-rich red giant stars can have CO absorptions that are as strong as the red supergiants found in starbursts; in other words, cold giants and slightly warmer supergiants can have equally strong CO line-strengths \citep{origlia1993, oliva1995}. This hampers the interpretation of CO absorptions in galaxies alone, in the absence of independent measurement of the stellar temperature. 

ETGs are known to host old stellar populations with little contribution, if any, from recently-formed stellar populations. Indeed, since~\citet{frogel1980}, the CO (2.3 $\mu$m) absorption has been used to possibly infer the presence of young stars (red giants and supergiants) in ETGs. In particular, \citet{mobasher2000} found that the CO line-strength is stronger for ellipticals in the outskirts of the Coma cluster than in the core, interpreting this as an evidence for younger populations in galaxies inhabiting low-density environments (see also~\citealt{james1999}). \citet{mobasher1996} and \citet{marmol2009} also interpreted the observed higher value of the $2.3~\mu$m CO line-strength of galaxies in the field with respect to those in denser environments of clusters as due to relatively more recent episodes of star formation in field galaxies. Unfortunately, most of these analyses has hitherto  been based on a direct comparison of CO indices in galaxies to those for stellar spectra, with no detailed comparison to predictions of SPS models. Recently, \citet{baldwin2018} measured the $2.3~\mu$m CO line-strength for 12 nearby ETGs, comparing to predictions from different sets of SPS models. They found that all models systematically underpredict the strength of the \textit{K}-band CO.

While the CO bandhead in the \textit{K} band has been extensively analyzed in the literature, no much effort has been done so far to study other NIR CO lines, that are prominent in galaxy spectra, especially in the \textit{H} band. This is because low-temperature and heavily obscured stars are brighter in the \textit{K} than the \textit{H} band, and, perhaps more importantly, severe contamination of the \textit{H} band spectral range from sky emission lines has prevented its exploitation in the work of the late 1900s. However, nowadays, thanks to the high-resolution of NIR spectrographs and new sky-subtraction techniques, the \textit{H}-band is fully accessible to detailed stellar population studies. 

CO absorptions in the \textit{K} band arise from transitions between two vibrational states with a difference ($\Delta\nu$) in quantum number $\nu$ of 2, whilst the $\Delta\nu$ of CO absorptions in the \textit{H} band is equal to 3 (see table 10 of \citet{rayner2009}). This results in a lower ($\sim$2 orders of magnitude) optical depth of CO lines in the \textit{H} band than the \textit{K} band and therefore, the CO lines in the \textit{H} band saturate in cooler stars than those in \textit{K} band. Hence the strength of the CO lines in the \textit{H} and \textit{K} band are expected to behave differently with spectral type and luminosity \citep{origlia1993}. Indeed, performing a simultaneous analysis of different features arising from the same chemical species is of paramount importance, as it helps in breaking degeneracies among relevant stellar population parameters (e.g.~\citealt{conroy2012a}). The only effort in this direction has been done by~\citet{riffel2019}, who have analyzed CO line-strengths in both the \textit{H} and \textit{K} band, for a sample of nearby ETGs, covering a range of galaxy mass, as well as star-forming galaxies. They found that while some CO lines are matched by the models, others seem to exhibit a significant disagreement. 

In this paper, we perform a detailed analysis of a whole battery of CO absorption features that are found in the NIR spectra of ETGs, focusing on a homogeneous, high-quality, sample of very massive nearby galaxies, with a velocity dispersion of $\sim300$~\kms\ (i.e. the high-mass end of the galaxy population), as well as other galaxy samples collected from previous works. We show that, indeed, all CO features, besides the well-studied $2.3~\mu$m CO bandhead, are systematically underestimated by the models for the very massive ETGs. We scrutinize several possible explanations of this ``CO mismatch'' problem, including the effect of a non-universal IMF, a contribution from young and intermediate-age populations, high-metallicity stars, as well as the effect of non-solar abundance ratios. We also present ad-hoc empirical SPS models that might help to solve the problem, by taking advantage of the scatter of stars in the available stellar libraries used to construct the models.

The paper is organised as follows: In Sections~\ref{sec:samples} and \ref{sec:libraries_models}, we describe our samples of ETGs, as well as the stellar libraries and synthesis models used in the present work.  In Section~\ref{sec:spectral_indices} we show the CO mismatch problem for NIR absorption features, by comparing models  and observations. Different experiments are presented in Section~\ref{sec:abundance_agb} in order to search for a possible solution of the CO puzzle. Our empirical approach to address the tension is explained in Section~\ref{sec:empirical_approach}. Section~\ref{sec:discussion} provides a discussion of our results. The overall conclusions are summarized in Section~\ref{sec:conclusions}.

\section{Samples} \label{sec:samples}

We used different samples of ETGs drawn from the literature. Our main galaxy sample is that of \citet{labarbera2019} (hereafter LB19), consisting of exquisite-quality, high-resolution, optical and NIR spectra for a sample of very massive ETGs at z $\sim0$, collected with the X-SHOOTER spectrograph \citep{vernet2011} at the ESO Very Large Telescope (VLT). Other samples of ETGs were used wherever the quality of data and wavelength coverage were suitable for our analysis.  Although these samples are far from being homogeneous (as they were observed with different instruments), they encompass a wide range in galaxy stellar mass and stellar population parameters, allowing for a comprehensive study of NIR CO features. In particular, we have included two samples of ETGs residing in varying environments (see below), namely in the field \citep{marmol2009}, and in the Fornax cluster \citep{silva2008}, whose \textit{K}-band spectra were obtained with the same observational setup. These two samples allow us to explore the effect of the environment on the CO strengths.

The main properties of our galaxy samples are summarized as follows: 

\begin{itemize}
    \item\citet{labarbera2019} (hereafter XSGs): The seven massive ETGs of this sample are at redshift $z\sim0.05$, and span a velocity dispersion ($\sigma$) range of $\sim 300 - 360$~\kms. Five galaxies are centrals of galaxy groups, while two systems are satellites (see LB19 for details). The galaxies were observed using the X-SHOOTER three-arm echelle spectrograph, mounted on UT2 at the ESO VLT. The wavelength coverage of the spectra in the NIR arm ranges from $9800$ to $25000$~\AA \space with a final resolution of $\sim5500$\,(FWHM). The spectra used for the present work were extracted for all galaxies within an aperture of radius $1.5$", and have a high signal-to-noise ratio (SNR), above $170$~\AA$^{-1}$. In addition to individual spectra, we also made use of a stacked spectrum in our analysis,  to characterize the average behaviour of the sample. Using optical (H${\rm \beta o}$, H$_{\rm \gamma F}$, TiO1, TiO2$_{\rm SDSS}$, aTiO, Mg4780, [MgFe]', NaD, \ion{Na}{i}8190) and NIR (\ion{Na}{i}1.14, \ion{Na}{i}2.21) spectral indices, LB19 showed that stellar populations in the center of these galaxies are old ($\sim11$~Gyr), metal-rich ($\sim+0.26$~dex), enhanced in $\alpha$-elements (\aFe$\sim+0.4$~dex) and with bottom-heavy IMF (with logarithmic slope $\Gamma_{b}\sim3$ for the upper segment of a low-mass tappered IMF, often regarded as "bimodal").
    
    \item\citet{francois2019}: The twelve nearby (z < $0.016$) and bright (B$_{t}\sim 11-13$~mag) galaxies of this sample span $\sim 35 - 335$~\kms \space in velocity dispersion and are distributed along the Hubble sequence from ellipticals to spirals. They have been observed with the X-SHOOTER spectrograph at the VLT. NIR spectra were obtained with a $1.2$" slit, providing a resolving power of $4300$. The one-dimensional spectra were extracted by sampling the same spatial region for all galaxies. Using optical indices (<Fe>, [MgFe]', Mg$_{2}$, Mg$_{b}$, and H$_{\rm \beta}$), \citet{francois2019} showed that stellar populations in the center of these galaxies span a wide range of values in age and metallicity ( $0.8 \leq \rm age \leq 15$ and $-0.4 \leq \rm \Zh \leq 0.5$). 
    
    \item\citet{baldwin2018}: They obtained \textit{JHK}-band spectra for twelve nearby ETGs from the ATLAS$\rm^{3D}$ sample \citep{cappellari2011}, with a velocity dispersion range of $\sigma = 80 - 120$~\kms, using GNIRS, the Gemini Near-Infrared Spectrograph at the 8m Gemini North telescope in Hawaii. The galaxies span a broad range in age from $1$ to $15$~Gyr (optically-derived SSP-equivalent ages) at approximately solar metallicity. They used a $0.3$" slit, with a spectral resolution of R $\sim 1700$. One-dimensional spectra were extracted within an aperture of $\pm\frac{1}{8}$R$_{\rm eff}$ except for one galaxy (whose spectrum was extracted within $\pm\frac{1}{12}$R$_{\rm eff}$). The SNR of the spectra is in the range $50-200$.
    
    \item\citet{marmol2009}: They observed twelve ETGs in the field in the velocity dispersion range $59$~\kms \space < $\sigma$ < $305$~\kms \space with the ISAAC NIR imaging spectrometer, mounted on UT1 at the VLT. NIR spectra were obtained with a $1$" slit, providing a resolving power of $7.1$~\AA \space at $2.3~\mu$m. The wavelength coverage of the spectra is short (from $2.20$ to $2.29~\mu$m), including \ion{Na}{i}, \ion{Fe}{i}, \ion{Ca}{i} and the first CO absorptions at the red end of the \textit{K} band (see below). They extracted galaxy spectra within a radius corresponding to $\frac{1}{8}$R$_{\rm eff}$. The ages of these galaxies were determined by \citet{sanchez2006b}, using the H$_{\rm\beta}$ index and a preliminary version of the SPS models of \citet{vazdekis2010}.
    
    \item\citet{silva2008}: This sample consists of eight ETGs in the Fornax cluster with $\sigma = 70 - 360$~\kms, all observed  with the ISAAC NIR imaging spectrometer. A $1$" slit was used during the observations, giving a spectral resolution of R $\approx$ $2900$ ($7.7$~\AA\ FWHM). The central spectra were extracted within a radius corresponding to $\frac{1}{8}$R$_{\rm eff}$ with a SNR of $48 - 280$, and covering a wavelength range of $2.12 - 2.37~\mu$m. The ages of these galaxies are drawn from \citet{kuntschner1998}, and were computed with \citet{worthey1994} SPS models.
    
\end{itemize}

Note that, for all spectra, we measured CO spectral indices (see Sec.~\ref{sec:spectral_indices}), when they lie within the available spectral range and are considered to be safe for the stellar population analysis, according to the criteria given in \citet{eftekhari2021} (see their section~4.8).

\section{Stellar libraries and stellar population models}\label{sec:libraries_models}

We compare observed CO index strengths with predictions of various SPS models which differ in a number of ingredients, such as the adopted isochrones, stellar libraries, IMF shape, age and metallicity coverage, as well as the prescription to implement the AGB phase. The latter is one of the most important aspect when studying the NIR spectral range, as it is the main source of differences seen among models. We also compare CO indices observed in our galaxy samples with individual stars, from theoretical and empirical stellar libraries. The main features of the models and libraries are summarized below.

\subsection{Stellar population models}\label{sec:models}

\begin{itemize}
    \item E-MILES: We used two model versions: base E-MILES \citep{vazdekis2016} and $\alpha$-enhanced E-MILES (an updated version of Na-enhanced models described in \citealt{labarbera2017}). The models are available for two sets of isochrones; BaSTI \citep{pietrinferni2004} and Padova00 \citep{girardi2000}. These isochrones provide templates for wide range of ages, from $1-14$~Gyr in BaSTI and $1-17.78$~Gyr for Padova00, and metallicities, from $-0.35$ to $+0.26$~dex in BaSTI and $-0.4$ to $+0.22$~dex in Padova00. The base model is computed for different IMF shapes- Kroupa universal, revised Kroupa, Chabrier, unimodal and bimodal- while the $\alpha$-enhanced model is only available for bimodal IMF distributions (see \citealt{vazdekis1996, vazdekis1997} for a description of different IMF parametrizations). E-MILES models cover a wide range in wavelength from ultraviolet ($1680$~\AA) to infrared ($50000$~\AA) and are based on NGSL stellar library \citep{gregg2006} in the ultraviolet, MILES \citep{sanchez2006a}, Indo-US \citep{valdes2004} and CaT \citep{cenarro2001} stellar libraries in the optical and 180 stars of IRTF stellar library (see below) in the infrared. For the current study, we utilised the BaSTI-based model predictions with bimodal IMF shape with slopes $\Gamma_{b}=1.3$ (representative of the Kroupa-like IMF) and $\Gamma_{b}=3.0$ (representative of a bottom-heavy IMF), and two metallicity values, around solar ($+0.06$) and metal-rich ($+0.26$). Note that the BaSTI models have cooler temperatures for low-mass stars and use simple synthetic prescriptions to include the AGB regime.
     
    \item Conroy et al.: We used two model versions: \citet{conroy2012a} (hereafter CvD12) and \citet{conroy2018} (hereafter C18).  The CvD12 models rely on 91 stars from the IRTF stellar library in the NIR,  using different isochrones to cover different phases of stellar evolution, from the hydrogen burning limit to the end of the AGB phase, namely the Dartmouth isochrones \citep{dotter2008}, Padova isochrones  \citep{marigo2008} and Lyon isochrones \citep{chabrier1997, baraffe1998}.  The publicly-available models of CvD12 cover either non-solar abundance templates at $13.5$~Gyr, or younger ages (from $3$ to $13.5$~Gyr) at solar abundance ratios and metallicity. The $13.5$ Gyr model of solar metallicity and solar abundance is available for a Salpeter IMF, Chabrier IMF, two bottom-heavy IMFs with logarithmic slopes of x=$3$ and x=$3.5$, and a bottom-light IMF. In this paper, we utilised C-, O-, and $\alpha$-enhanced models of age $13.5$~Gyr in addition to solar metallicity models of ages $3$ to $13.5$~Gyr with a Chabrier IMF. The updated version of CvD12 models is based on the MIST isochrones \citep{choi2016} and utilises the Extended IRTF Stellar Library \citep{villaume2017} that includes continuous wavelength coverage from 0.35 to 2.4~$\mu$m. C18 models cover a wide range in metallicity ($-1.5 \lesssim {\rm[Fe/H]} \lesssim 0.3$) and include new metallicity- and age-dependent response functions for 18 elements. In this paper, we used C18 models with overall metallicity [Z/H] of 0.0 and 0.2~dex, respectively, with a Kroupa-like IMF, for ages between 1 to 13~Gyr. We also used C-enhanced models ([C/Fe] = +0.15) of age 13~Gyr, solar metallicity and a Kroupa-like IMF.
    
    \item\citet{maraston2005} (hereafter M05): This model is based on the fuel consumption theorem of \citet{renzini1981} for the post main sequence stages of the stellar evolution (in particular TP-AGBs) and makes use of two stellar libraries, the BaSel theoretical library \citep{lejeune1998} in the optical and NIR  and the \citet{lancon2000} empirical library of TP-AGB stars. The models are available for Salpeter and Kroupa IMF and for two horizontal branch (HB) morphologies: blue and red. They cover a wide range in metallicity (from $-2.35$ to $+0.67$~dex) and age (from $10^{3}$ yr to $15$~Gyr). For our analysis we have used models with solar metallicity, Kroupa IMF and ages between $1$ and $14$ Gyr, with red HB morphology.
    
\end{itemize}

\subsection{Stellar libraries}\label{sec:libraries}

\begin{itemize}
    \item IRTF \citep{cushing2005, rayner2009}: This is an empirical spectral library of $210$ cool stars covering the \textit{J}, \textit{H}, and \textit{K} bands ($0.8-5~\mu$m) at a resolving power of R $\sim2000$. The library includes late-type stars, AGB, carbon and S stars, mostly with solar metallicity\footnote{Note that, although the IRTF library has been recently extended to a wider range in metallicity by \citet{villaume2017}, in the present paper we used the original version of the IRTF library, as this library is actually used to build up our reference SSP models (E-MILES).    Moreover, the extended library shows a significant improvement mostly in the low-metallicity regime (see figure~1 of \citet{villaume2017}), which is not relevant for our samples of massive ETGs.}, providing absolute flux calibrated spectra. For this study, we have used a subsample of $180$ IRTF stars that are also used to construct E-MILES models in the NIR. 
    
    \item Theoretical stars from \citet{knowles19_Thesis}: We included a small set of theoretical stellar models, generated using the same method presented in \cite{knowles21}. In summary, these models are computed using ATLAS9 (\citealt{kurucz1993}), with publicly available\footnote{\url{http://research.iac.es/proyecto/ATLAS-APOGEE//}} opacity distribution functions described in \citealt{mezaros2012}. We used the 1{\small D} and LTE mode of ASS$\epsilon$T (Advanced Spectrum SynthEsis Tool, \citealt{koesterke2009}) with input ATLAS9 atmospheres to produce fully consistent synthetic spectra at air wavelengths, with abundances varied in the same way in both model atmosphere and spectral synthesis components. The models here adopt \citet{asplund2005} solar abundances and a microturbulent velocity of $2$~\kms. We direct interested readers to \cite{knowles19_Thesis} and \cite{knowles21} for further details. The star models in this work have effective temperatures of $3500$, $3750$ and $4000$~{\small K}, for \logg=$1.5$, \Mh=\aM=$0.0$ and two different carbon abundances; scaled-solar (\CFe=$0.0$) and enhanced (\CFe=$0.25$). \Mh\ here is defined as a scaled-metallicity in which all metals, apart from the $\alpha$-elements and carbon if they are non-solar, are scaled by the same factor from the solar mixture (e.g. \Mh=$0.2$=\Feh=\liH). This definition results in \aM=\aFe\ and \cM=\CFe. The models are generated specifically for this work, with a wavelength range of $1675.1$-$24002.1$~\AA\ and a resolution of R $\approx100000$. 
    
    \item APOGEE \citep{majewski2017}: APOGEE is an \textit{H}-band ($1.59 - 1.69~\mu$m) spectral library of stars in the Milky Way from the SDSS with a resolving power of $\sim22500$. The $\rm16^{th}$ data release provides reduced and pipeline-derived stellar parameters and elemental abundances (ASPCAP; \citealt{garcia2016}) for more than $430,000$ stars. For our study, in the ASPCAP catalog, we made a selection of a subset of stars\footnote{We removed stars with bits STAR\_BAD, TEFF\_BAD, LOGG\_BAD, and COLORTE\_WARN set in the APOGEE\_ASPCAPFLAG bitmask and also those with bits BRIGHT\_NEIGHBOR, VERY\_BRIGHT\_NEIGHBOR, PERSIST\_HIGH, PERSIST\_MED, PERSIST\_LOW, SUSPECT\_RV\_COMBINATION, and SUSPECT\_BROAD\_LINES set in the APOGEE\_STARFLAG bit-mask.}. We also removed stars with SNR < $100$ (per pixel) and those that have a radial velocity scatter greater than $1.5$~\kms, and a radial velocity error greater than $3$~\kms.
    
\end{itemize}

\section{CO spectral indices}\label{sec:spectral_indices}

As described in Sec.~\ref{sec:intro}, fitting SPS models to CO absorption features of galaxies in the \textit{K} band has proven to be challenging. To gain further insights into the problem, we consider here not only the already studied CO feature at $\sim2.30~\mu$m, but also other CO features at $2.32~\mu$m and $2.35~\mu$m. These are the first-overtone CO bandheads at the red end of the \textit{K} band; their depth increases with decreasing stellar temperature \citep{kleinmann1986, rayner2009} and luminosity \citep{origlia1993} and becomes progressively weaker with decreasing metallicity \citep{frogel1975, aaronson1978, doyon1994, davidge2018, davidge2020}, being stronger in red giant and supergiant stars than in dwarf stars\footnote{The first overtone bands of $^{12}$CO show bandheads at 22929.03, 23220.50, 23518.16, and 23822.97~\AA\ (in air) and those of $^{13}$CO at 23441.84, 23732.94, and 24030.49~\AA. The feature at 2.35~\AA\ is dominated by the $^{12}$CO bandhead at 23518.16~\AA\ but disturbed by the $^{13}$CO line at 23441.84~\AA. The strong saturation of the $^{12}$CO feature makes the $^{13}$CO line clearly visible even for low $^{12}$CO/$^{13}$CO abundance ratios (see the shoulder on the blue-ward of the central feature in panel CO2.35 of Fig.~\ref{fig:fig1}). Consequently, the first overtone bands of CO can be used to estimate the $^{12}$CO/$^{13}$CO isotope ratio providing, in principle, important clues about stellar evolution and nucleosynthesis.}. In addition, we also include in our analysis six second-overtone CO bandheads in the \textit{H} band, namely at $1.56$, $1.58$, $1.60$, $1.64$, $1.66$, and $1.68~\mu$m. In the following, we first show observed and model spectra around each CO absorption (Sec.~\ref{sec:obs_vs_mod}), and then we discuss observed and model line-strengths in the \textit{K} and \textit{H} bands (see Secs.~\ref{sec:K_band} and~\ref{sec:H_band}), respectively. 

\subsection{CO indices: observed vs model spectra} \label{sec:obs_vs_mod}

\begin{figure*}
\centering
	\includegraphics[width=.97\linewidth]{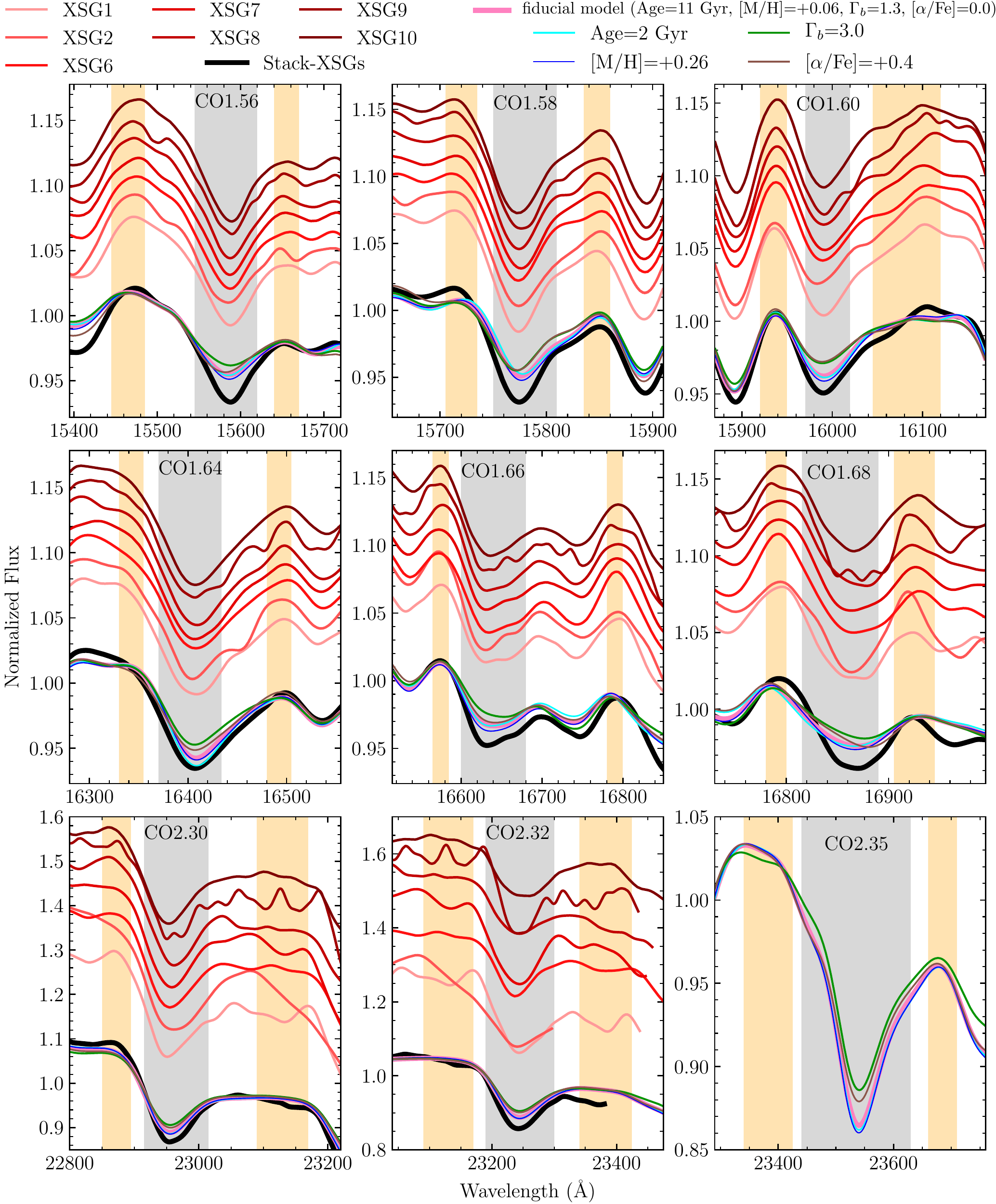}
    \caption{Spectra of XSGs and E-MILES models in the regions of CO features: individual XSG spectra (light to dark red); the stacked spectrum of XSGs (black); a ``fiducial model'' with solar abundance and metallicity, an age of $11$~Gyr, and MW-like IMF (pink); a model with Age = $2$~Gyr (cyan); a model with \Mh\ = $+0.26$ (violet), a model with a bottom-heavy IMF of $\Gamma_{b} = 3.0$ (green); and a model with \aFe\ = $+0.4$ (brown). All these models have the same stellar population parameters as the fiducial model except for a given parameter (see labels on the top). The model and observed spectra have been convolved to a common resolution of $\sigma = 360$~\kms, as this is the highest velocity dispersion in the sample of \citet{labarbera2019}. The central bandpasses of CO indices, as well as the blue and red pseudo-continua, are from \citet{eftekhari2021} and are shown as grey and orange areas, respectively. All spectra have been normalised to the mean flux within pseudo-continua bands. The XSG individual spectra have also been arbitrarily shifted to display galaxy-to-galaxy variations in the depth of the COs. Note that the spectral range of XSGs does not cover the CO2.35 index but this index is included in the figure as it is used for other galaxy samples. Remarkably, for all CO features, galaxies show stronger absorption than the models, regardless of the adopted model parameters.
    }
    \label{fig:fig1}
\end{figure*}

In Fig.~\ref{fig:fig1}, we show the spectra of XSGs from LB19, around CO absorptions from \textit{H} through \textit{K} band, and compare them with model spectra. From light to dark, the shifted red spectra correspond to the individual XSGs, while the median-stacked spectrum is shown in black. The wavelength definitions of CO indices, from \citet{eftekhari2021}, are shown with shaded grey and orange areas corresponding to indices bandpass and pseudo-continua bands, respectively. In the same figure, we also show a fiducial model (pink) corresponding to an E-MILES simple stellar population (SSP) with an age of $11$~Gyr (mean age of the XSGs as derived from optical indices), solar chemical abundance pattern, and a bimodal IMF of logarithmic slope $1.3$ (corresponding to a MW-like IMF). Note that although the spectra of XSGs do not cover the CO index at $2.35~\mu$m, we also show this index as it is covered by other galaxy samples in our analysis (see Fig.~\ref{fig:fig2}). For clarity and ease of comparison, normalised spectra of XSGs are shifted upwards while the stacked spectrum and model spectra are only normalised to the mean flux within pseudo-continua bands. Clearly, XSG2 is the one with the largest scatter amongst the spectra, likely because of the lower quality of the data for this galaxy, that has been observed with a different observational setup (see LB19 for details). A clear mismatch between observations and models is seen in all panels. The CO indices of the XSGs are much stronger than those of the fiducial SSP model. 

As a first step, we investigate whether a young population, a non-solar metallicity, a dwarf-rich stellar population, and/or a non-solar chemical abundance pattern could significantly affect the CO lines, and explain the deep CO absorptions in the data:

\begin{description}
\item[-] As mentioned in Sec.~\ref{sec:intro}, the disagreement between \textit{K}-band CO observations and models has been attributed to the presence of intermediate-age stellar components, dominated by stars in the AGB evolutionary phase. To test this scenario, in Fig.~\ref{fig:fig1}, we show an E-MILES SSP with the same parameters as the fiducial model but with an age of $2$~Gyr (see cyan curves). Except for the CO1.64 index, a variation in age does not significantly change the depth of CO absorption features. We assess this issue, in more detail, in Sec.~\ref{sec:AGB}. 
\item[-] Since XSGs have metal-rich stellar populations, as shown by the analysis of optical spectral indices (see LB19), in the figure we also show the effect of increasing metallicity, with an SSP having the same parameters as the fiducial model but \Mh = $+0.26$ (violet). According to E-MILES models, the increase in metallicity does not significantly affect the depth of (all) CO features. 
\item[-] The strong CO absorptions can not be explained by IMF variations either; an SSP model with the same fiducial model parameters but with a steeper IMF slope (green) has shallower CO absorptions, hence worsening the fitting. This was first pointed out by \citet{faber1972}, who showed that an increase in the number of dwarf stars drives the CO index at $22800$~\AA\ to unacceptably low values. Figure~\ref{fig:fig1} shows that, indeed, all CO features exhibit a similar behaviour.
\item[-]In Fig.~\ref{fig:fig1}, we also investigate non-solar \aFe\ abundance effects. An SSP that differs from the fiducial model only in $\alpha$-enhancement is shown with brown colour. As for the IMF, the enhancement in $\alpha$ weakens the strength of CO absorptions. Notice that a different result seems to hold for A-LIST SPS models \citep{ashok2021}, which suggest an increase of CO strength with \aFe\ (see their figure 6.a).
\end{description}

\subsection{CO line-strengths in \textit{K} band} \label{sec:K_band}

Figure~\ref{fig:fig2} shows a quantitative comparison 
of line-strengths of CO indices in \textit{K} band between data and different SPS models (see Sec.~\ref{sec:models}), i.e. E-MILES (solid/dotted pink and purple lines), CvD12 (dashed violet line), C18 (solid and dotted violet lines), and M05 (solid black line) models. We plot model line-strengths of the COs as a function of age, while for galaxies we plot observed line-strengths as a function of the age, as  estimated from previous works (see Sec.~\ref{sec:samples}).  For each index, the measurements on the individual XSG spectra are plotted  with open red circles, while the filled red circle corresponds to the measurement of their stacked spectrum. Individual galaxies from \citet{francois2019}, \citet{baldwin2018}, \citet{marmol2009}, and \citet{silva2008} samples (hereafter B18, F19, M09, and S08, respectively) are shown with open lime squares, cyan triangles, blue pentagons, and orange diamonds, respectively. For each index, the median CO line-strength of each sample is shown with a filled symbol of the same colour. The indices were measured after smoothing all spectra to a common velocity dispersion of $\sigma = 360$~\kms. 

\subsubsection{D$_{\rm CO}$ vs. age}
In the upper-left panel of Fig.~\ref{fig:fig2}, we consider the D$_{\rm CO}$ index, i.e. the definition of the first CO bandhead in \textit{K} band from \citet{marmol2008}. This index is defined with two blue pseudo-continua and the absorption bandpass (see \citealt{marmol2008} for details). In the same panel, we also included measurements for the spectra of S08 and M09  (open/filled orange diamonds and blue pentagons). For ages greater than $3$~Gyr, different models show similar trends, with D$_{\rm CO}$ showing no significant variation with age. Only for ages younger than $3$~Gyr, there is a significant difference between E-MILES and M05 models since the contribution of AGB stars are more emphasized in the young populations of M05. We further discuss this issue in Sec.~\ref{sec:AGB}. All models underpredict the median value of D$_{\rm CO}$ for the samples except for that of F19. However, since the scatter of D$_{\rm CO}$ in this sample is far larger than that for the other samples, no firm conclusions can be drawn. In general, Fig.~\ref{fig:fig2} shows that the models can barely reproduce only the galaxies with the smallest D$_{\rm CO}$ values. For instance, two galaxies of B18 at $13$~Gyr are well matched with E-MILES and M05 models,  and the same applies to two galaxies in the S08 sample with the weakest CO absorption, that are well matched with E-MILES models (see solid pink line and the orange diamonds for an age of $\sim 8$~Gyr). Since the M05 model differs significantly from E-MILES model in the predictions for young populations, it can match the youngest galaxies of M09 and B18. Overall, for  the D$_{\rm CO}$ index, the mismatch between observations and models applies to all models. C18 models predict the lowest values for D$_{\rm CO}$ and cannot match any data points.

\subsubsection{D$_{\rm CO}$ vs. metallicity}
Another parameter that can be considered is the variation of metallicity among our galaxies, that span a wide range, from \Mh = $-0.4$ to \Mh = $+0.5$~dex. We investigate the effect of metallicity by showing the predictions of E-MILES models with a MW-like IMF ($\Gamma_{b}=1.3$) and total metallicity of $+0.26$~dex (dotted-pink line), and predictions of C18 models with a Kroupa IMF and [Z/H] = 0.2~dex (dotted-violet line). Hence, the effect of metallicity can be seen by comparing dotted and solid lines in Fig.~\ref{fig:fig2}.  The figure shows that the discrepancy between the dotted-pink line and the filled blue pentagon and red circle (orange diamond) is almost 8 (2) times larger than the increase in D$_{\rm CO}$ caused by variations in \Mh\ from $+0.06$ to $+0.26$ dex.
However, it is worth noticing that the IRTF stellar library, that is used to construct E-MILES models in the NIR, consists of stars in the solar neighbourhood, which are unavoidably biased towards solar metallicity. In fact, according to \citet{rock2015}, the quality of E-MILES models decreases at supersolar metallicities. Also,  we cannot exclude a non-linear behaviour of CO absorptions in the very high-metallicity regime. Therefore, we looked at C18 models, which are based on the Extended IRTF Stellar Library with better coverage in metallicity (although this advantage mainly applies to the metal-poor regime). Comparing the difference between dotted and solid pink lines with the difference between dotted and solid violet lines shows that the effect of metallicty in C18 models is larger than the one in E-MILES models but it is not enough to match the high values of XSGs or M09 data. Also, notice that although C18 models are based on a stellar library with better coverage in metallicity, they predict the lowest values for D$_{\rm CO}$, hampering the discrepancy to the observed data points. We conclude that the mismatch between data and models cannot be explained with metallicity variations alone.

\subsubsection{D$_{\rm CO}$ vs. IMF}
\label{sec:DCO_IMF}
We also investigated the effects of a bottom-heavy IMF ($\Gamma_{b}=3.0$) on the CO features as shown by solid- and dotted-purple lines for solar and metal-rich populations, respectively. The IMF slope of XSGs has been determined by LB19, using a combination of optical and NIR (Na) indices, finding that all galaxies have a bottom-heavy IMF in the centre. However, Fig.~\ref{fig:fig2} shows that a dwarf-rich population does significantly increase the discrepancy between models and observed CO indices. While this result seems to be consistent with \citet{alton2018}, who claimed a MW-like IMF for massive galaxies in their sample, based on \textit{J}- and \textit{K}-band spectral indices (including two CO bandheads in \textit{K} band), it has to be noted that for a MW-like IMF the models do not match the observations either. In other words, any claim from CO lines on the IMF should be taken with caution.

\subsubsection{D$_{\rm CO}$ vs. environment}
The  samples of S08 and M09 allow us to assess the effect of galactic environment on the strength of D$_{\rm CO}$, as galaxies in the former sample reside in a high-density environment (Fornax cluster) compared to the latter, which consists of field galaxies. 

It is noteworthy to mention that within 20~Mpc, the Fornax cluster is the closest and second most massive galaxy cluster after Virgo. It has a virial mass of 10$^{13}$\ms \space \citep{drinkwater2001} and while most of its bright members are ETGs, mainly located in its core \citep{grillmair1994}, its mass assembly is still ongoing \citep{scharf2005}, and therefore it is not fully virialized. Fornax cluster is an evolved, yet active environment, as well as a rich reservoir for studying the evolution of galaxies in a cluster environment, particularly within its virial radius.

The two samples of S08 and M09 have been observed with the same telescope and observational setup (see Sec.~\ref{sec:samples}), allowing for a direct comparison. Note that these two samples are also comparable with respect to velocity dispersion (see Sec.~\ref{sec:samples}). As shown in the plot, M09 galaxies, located in the field, tend to have larger values of D$_{\rm CO}$ strength than S08 galaxies in the Fornax cluster. We see that the median value of these two samples cannot be matched with the current models. We can speculate that the origin of the difference between D$_{\rm CO}$ values of ETGs in low- and high-density environments might be due to a difference in the carbon abundance of field and cluster galaxies. Since star formation in dense environments takes place more rapidly than in isolated galaxies, carbon, which is expelled into the interstellar medium by dying stars of intermediate masses, cannot be incorporated in newer stellar generations. Therefore, the resulting stars in dense environments, like cluster galaxies, exhibit smaller carbon abundance with respect to their counterparts of similar mass in low-density environments. As the CO molecule has high binding energy, carbon mostly forms CO molecules. Thus CO indices in the field galaxies are stronger than cluster galaxies as was suggested by M09. Moreover, \citet{rock2017} found a dichotomy between the \ion{Na}{i}2.21 values of the ETGs in the S08 and M09 samples. Indeed, they showed that one possible driver of NaI2.21 might be \CFe\ abundance. 

\subsubsection{Other \textit{K}-band CO indices}
\begin{figure*}
\centering
	\includegraphics[width=\linewidth]{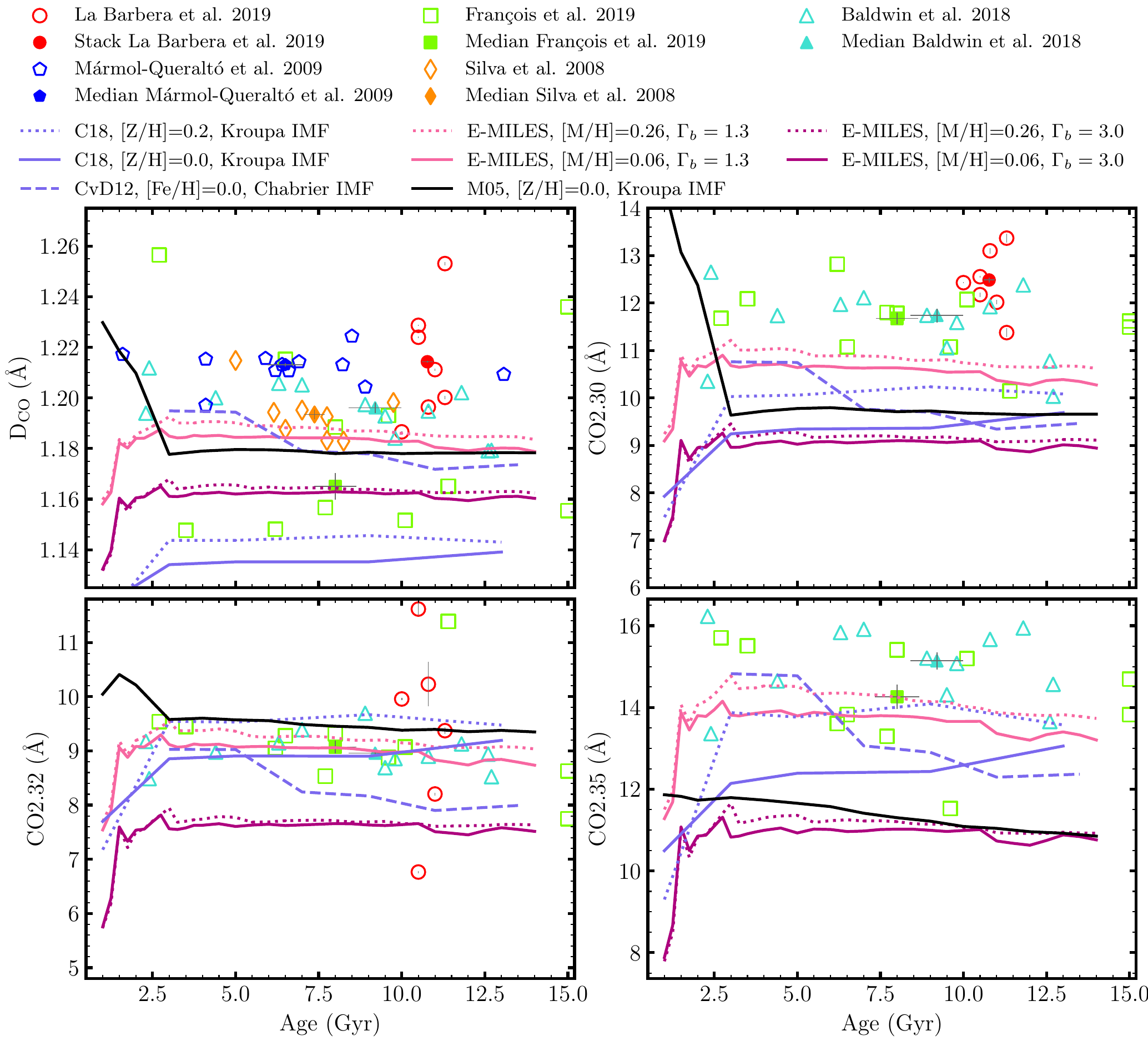}
    \caption{CO indices measured from different galaxy samples, with open (filled) lime squares, cyan triangles, blue pentagons, and orange diamonds corresponding to individual (median) values for galaxies from \citet{francois2019}, \citet{baldwin2018}, \citet{marmol2008}, and \citet{silva2008}, respectively (see labels on the top). Red open (filled) circles are line-strengths for individual spectra (median-stacked spectrum) of the XSGs \citep{labarbera2019}. Observed line-strengths are compared to CO indices measured on different SPS spectra: two sets of E-MILES models with solar metallicity and bimodal IMF of slopes $1.3$, corresponding to a MW-like IMF, and $3.0$, representative of the bottom-heavy IMF of massive galaxies (solid pink and purple lines, respectively); two sets of E-MILES models with super-solar metallicity and bimodal IMF slopes of $1.3$ and $3.0$ (dotted pink and purple lines); one set of CvD12 models with \Feh\ = $0$ and Chabrier IMF (dashed violet line); and one set of M05 models with solar metallicity and Kroupa-like IMF (solid black line). The indices are measured on data and models all corrected to the same velocity dispersion of 360~\kms. }
    \label{fig:fig2}
\end{figure*}

The upper-right panel of Fig.~\ref{fig:fig2}, shows  measurements for the same CO feature as for D$_{\rm CO}$, but using the index definition, named CO2.30, from \citet{eftekhari2021}. While D$_{\rm CO}$ measures the absorption at $\sim2.30~\mu$m as a generic discontinuity, defined as the ratio between the average fluxes in the pseudo-continua and in the absorption bands, the CO2.30 index follows a Lick-style definition, with a blue and red pseudo-continua and the absorption bandpass (see figure A3 in \citealt{eftekhari2021} for a comparison of the two definitions). Note that the red bandpass of CO2.30 is not covered by the S08 and M09 spectra and, therefore, these samples are not included in the upper-right panel of Fig.~\ref{fig:fig2}. For ages of 3 and $5$~Gyr, the predictions of CvD12 and E-MILES models for CO2.30 are very similar, while for older ages, CvD12 models are closer to the M05 predictions, leading to a larger mismatch. Also, C18 models with solar metallicity (solid violet line) have smaller difference with M05 models than E-MILES ones for ages older than 3 Gyr and their trend is very similar to E-MILES and M05 models, in contrast to CvD12. The difference between E-MILES, CvD12, C18 and M05 models for old ages (>$5$~Gyr) is quite significant, and comparable to the effect of varying the IMF slope (see pink and purple lines in the figure). However, similarly to D$_{\rm CO}$, all models fail to match the observed strong CO2.30 line-strengths, and changing \Mh\ of E-MILES models from $+0.06$ to $+0.26$, increases CO2.30 by only $\sim0.2$~\AA, while the discrepancy between the pink line and the filled lime square and cyan triangle (red circle) is $\sim1 (2)$~\AA. Moreover, although increasing the overall metallicity of C18 by 0.2~dex leads to an increase of $\sim1$~\AA, the C18 models predict the lowest values for CO2.30 and cannot match the data. Even by considering only the relative changes and adding the effect of metallicity predicted by C18 to the E-MILES models, the models are not still able to match the high value of the stacked spectrum of XSGs. Hence, using the CO2.30 index, we end up with the same conclusions as for D$_{\rm CO}$, i.e. an intrinsic offset exists between models and data, which is independent of the index definition.

In the lower panels of Fig.~\ref{fig:fig2}, we also show  measurements for other two CO bandheads in \textit{K} band, namely CO2.32 and CO2.35, which have been far less studied compared to the CO feature at $\sim2.30~\mu$m. Note that the XSGs spectra do not cover the wavelength limits of the CO2.35 index, and thus this sample is not included in the panel showing this index. Also, as can be seen in Fig.~\ref{fig:fig1}, the spectra of one XSG, and for the XSG stack, do not encompass the red bandpass of CO2.32. Hence, the corresponding measurements are not seen in the CO2.32 vs. age panel. Unlike CO2.30, for CO2.32, the M05 models predict the highest line-strengths among all SPS models with solar metallicity ($\sim0.5$~\AA\ higher than E-MILES and C18 models with a Kroupa IMF). E-MILES models with solar metallicity and MW-like IMF match well the median CO2.32 value of the F19 and B18 samples, and C18 models with solar metallicity are very close to F19 and match well to B18, while the scatter of XSGs is large, likely because the feature is at the edge of the available spectral range for these galaxies, making all models compatible with them. However, it should be noted that galaxies in these samples span a wide range in velocity dispersion, and at the highest $\sigma$, galaxies should be better described by a bottom-heavy IMF (see, e.g., \citealt{labarbera2013}). The latter is particularly true for the XSGs, with $\sigma \gtrsim 300$~\kms. It is expected that MW-like IMF models (pink lines) describe low-$\sigma$ galaxies, while bottom-heavy IMF models (purple lines) match high-$\sigma$ galaxies, in particular the XSGs (but see \citealt{alton2018}). However, predictions for a bottom-heavy IMF (purple lines) fall below the observed line-strengths for CO2.32, for all galaxies (but for one XSG). Thus, the mismatch between observations and models seems to be in place also for the CO2.32 index. Also, there is a closer similarity between the trend of C18 models and E-MILES and M05 models than CvD12 ones. Another interesting point is that C18 models with [Z/H] = 0.2~dex overpredict the mean CO2.32 line-strengths of F19 and B18 samples. Since the IMF of the most massive galaxies in F19 should be bottom-heavy, one might consider the effect of supersolar metallicity (from C18) and bottom-heavy IMF (from E-MILES) at the same time. In this case, the discrepancy between the C18 models and F19 sample gets even worse as the effect of IMF is larger than the effect of metallicity. The comparison for CO2.35 is similar to that for CO2.30, with solar and MW-like IMF models underpredicting the median values for F19 and B18 samples. However, the metal-rich E-MILES models (dotted-pink line) can reproduce the median value of the F19 galaxies (filled lime square), but as already pointed out, the most massive galaxies in this sample are expected to have a steeper IMF, while bottom-heavy models fall below all the observed data-points for CO2.35 (the bottom-heavy E-MILES model predicts a CO2.35 of $\sim11$~\AA, while the lowest value of CO2.35 in the F19 sample is $\sim11.5$~\AA). We conclude that, in general, the disagreement between observations and models is present in all the \textit{K}-band CO indices. 

\subsection{\textit{H}-band CO indices} \label{sec:H_band}

\begin{figure*}
\centering
	\includegraphics[width=\linewidth]{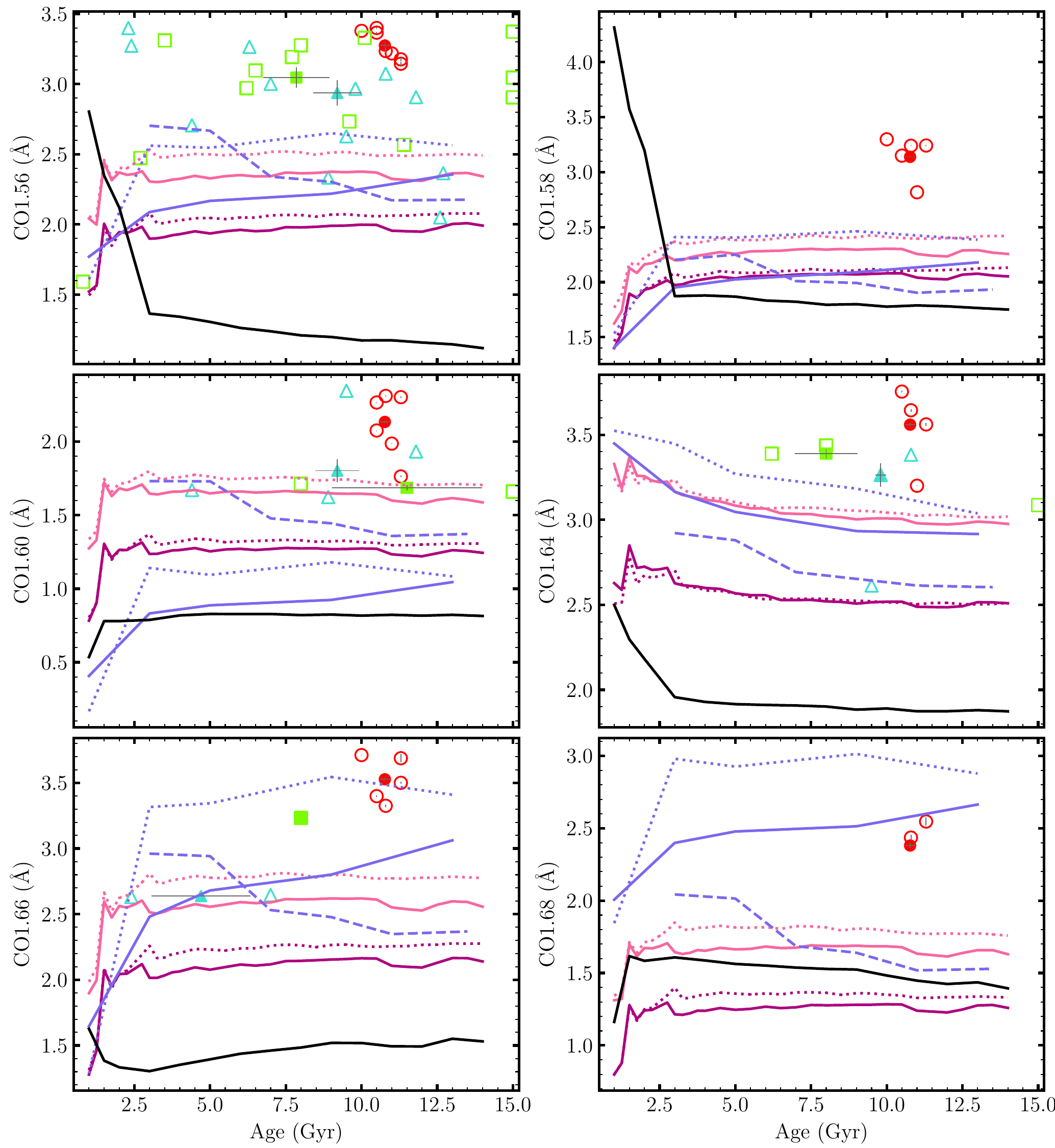}
    \caption{ The same as in Fig.~\ref{fig:fig2} but for \textit{H}-band CO indices.}
    \label{fig:fig3}
\end{figure*}

 In order to assess whether the mismatch of observed and model CO lines is intrinsic to the \textit{K} band, or whether it is a general issue in the NIR, we measured a whole battery of CO absorptions that populate the \textit{H}-band spectral range (see Fig.~\ref{fig:fig1}). Figure~\ref{fig:fig3} shows the same comparison as in Fig.~\ref{fig:fig2} but for the \textit{H}-band lines. Note that for CO1.58 and CO1.68, the spectra of F19 and B18 are severely contaminated by sky, and thus we do not show the corresponding line-strengths. For the same reason, only two XSGs are shown for CO1.68, and only a few galaxies are plotted in the panels for CO1.60, CO1.64, and CO1.66. 

 Remarkably, Fig.~\ref{fig:fig3} shows that (i) LB19 galaxies have lower scatter in all plots compared to the CO indices in \textit{K} band, most likely due to the high-quality of these data in \textit{H} band, and (ii) these very massive ETGs show very high CO values with respect to the model predictions for all indices. The discrepancy between models and observations for \textit{H}-band CO indices is similar to that found in \textit{K} band. In particular, the median stacked spectrum of the XSGs shows \textit{H}-band CO values $\sim1.3$ times larger than E-MILES models with MW-like IMF and solar metallicity. 
 
 For CO1.56, the offsets between the median values of F19 and B18 samples and the reference E-MILES model (pink line) are $\sim0.7$ and $\sim0.6$~\AA. E-MILES models can reproduce the median value of F19 and B18 galaxies for CO1.60 and CO1.66, respectively. However, these indices have been computed only for two galaxies in either samples, and thus we are not able to draw any firm conclusion. 
 
 Although the updated version of CvD12 models with solar metallicity (solid violet lines) tend to increase slightly for ages older than 3~Gyr (except for CO1.64), the behaviour of supersolar C18 models (dotted violet lines) is more similar to the E-MILES models and in case of CO1.58 it even matches with the E-MILES (dotted pink and violet line). For CO1.56 and CO1.64, C18 models with [Z/H] = 0.2~dex predict the highest values among all models but still they cannot match the median values of galaxies. The mean value of CO1.66 index in the B18 sample is well fitted by a solar metallicity C18 SSP and interestingly, a C18 SSP with supersolar metallicity matches the CO1.66 index value of stacked XSGs well. This result is at first glance surprising, however, one should bear in mind that XSGs have bottom-heavy IMFs and according to E-MILES models this makes the predictions of the CO values lower. In the CO1.68 panel, surprisingly, C18 models overpredict the line-strengths of the XSGs. In general, E-MILES and CvD12 models are more self-consistent than C18 and M05 models as their deviation with respect to data is similar for all CO indices. 

 In all panels of Fig.~\ref{fig:fig3}, the M05 models predict the lowest CO strengths, compared to other models. For CO1.56, CO1.58, and CO1.64, M05 models predict a strong increase at ages younger than $3$~Gyr, similar to what is found for CO2.30 in \textit{K} band, while for other indices the opposite behaviour is seen (e.g. CO1.60 and CO1.68). On the contrary, for all CO indices, the XSGs data show, consistently, line-strengths significantly above the model predictions for old ages (except for CO1.66 in which the supersolar metallicity C18 model matches the stacked spectrum and for CO1.66 in which C18 models overpredict the line-strengths of the XSGs). Again, this points against a scenario whereby the CO line-strengths are accounted for by young stellar populations with an AGB-enhanced contribution such as in M05 models. As for the \textit{K} band, CvD12 models show a trend for all CO indices to decrease with increasing age, while E-MILES models exhibit a nearly flat behaviour and C18 models a slightly increasing behaviour. For instance, in case of CO1.58, CO1.66, and CO1.68, for ages younger than $5$~Gyr, CvD12 models predict $\sim0.4$~\AA\ stronger line-strengths than E-MILES models (pink line),  while the two models agree for populations with an age of $\sim7$~Gyr. For older ages, CvD12 models always predict lower CO index values than E-MILES models. 

The effect of a bottom-heavy IMF in Fig.~\ref{fig:fig3} is shown by the purple lines, corresponding to E-MILES models with a bimodal IMF slope of $\rm \Gamma_b=3.0$. Similarly to the \textit{K} band, a bottom-heavy IMF leads to significantly  shallower CO line-strengths, than those for a standard stellar distribution. Note also that for most indices, the discrepancy between MW-like IMF models and the XSG stack (filled red circle) is larger than the variations due to a change in IMF slope. In the case of CO1.56 and CO1.58, the discrepancy is about twice the difference between models with different IMF. As already quoted in Sec.~\ref{sec:DCO_IMF}, we emphasize that while a bottom-heavy IMF hampers the offset between model and observations, we are not able to match the CO line-strengths with a standard IMF (except for CO1.66 with supersolar metallicity C18 model). In other words, the CO-strong feature should not be interpreted as evidence against a varying IMF in (massive) galaxies. 

From Figs~\ref{fig:fig2} and~\ref{fig:fig3}, a very consistent picture emerges. CO lines throughout the \textit{H} and \textit{K} bands are stronger than the predictions of all the considered state-of-the-art SPS models with varying age, metallicity, and IMF\footnote{At this point, we remind the reader that it is yet unclear why some indices, e.g. CO2.30, increase significantly for young populations in M05 models while they decrease in other models.}.  Indeed, in order to reconcile models and observations, additional stellar population parameters should be taken into account, as we detail in the following sections.

\section{Effects of varying other stellar population parameters}\label{sec:abundance_agb}

\begin{figure*}
\centering
	\includegraphics[width=\linewidth]{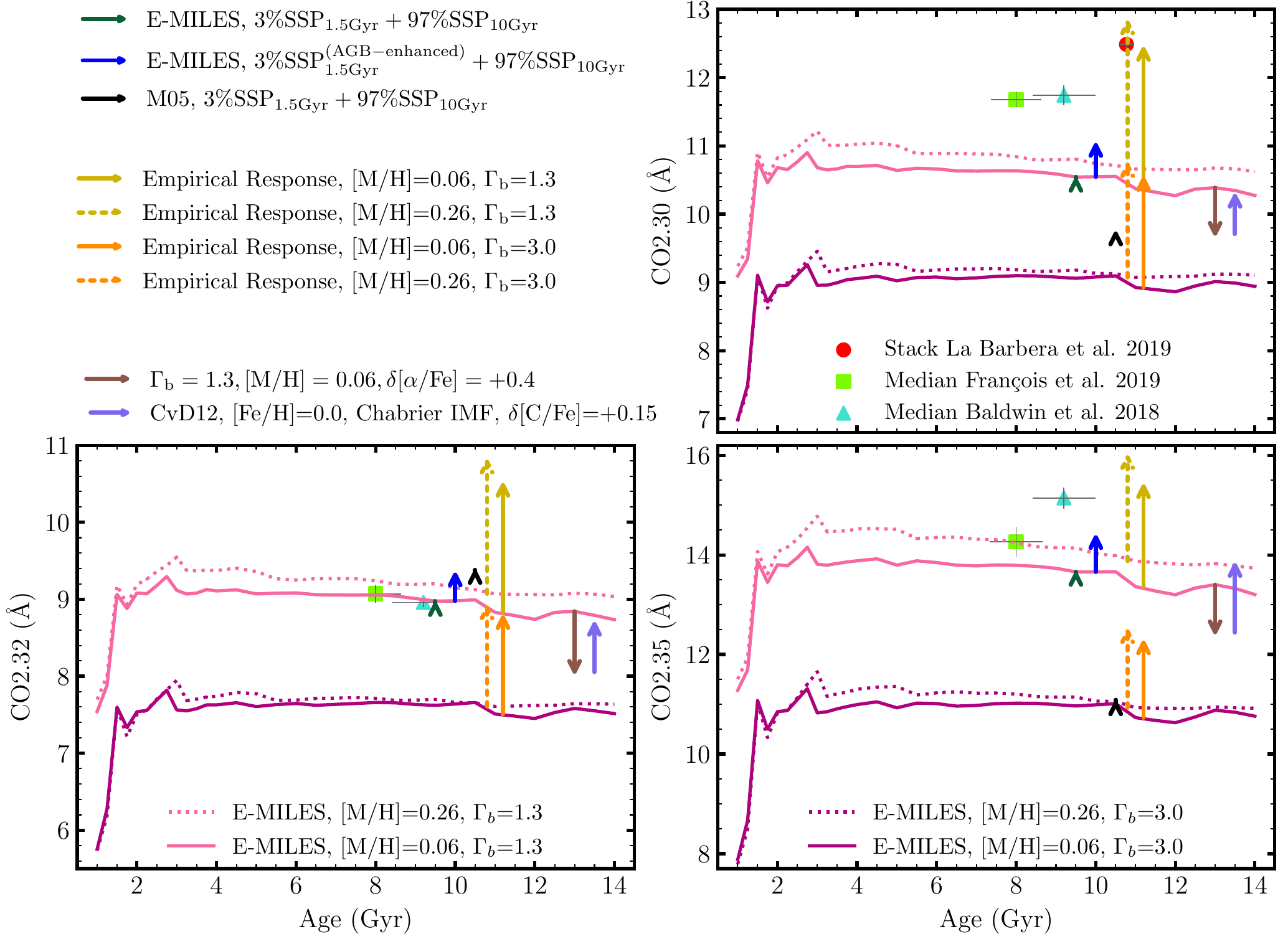}
    \caption{CO indices in \textit{K} band, as measured from different galaxy samples. Filled lime squares and cyan triangles correspond to median values for galaxies from F19 and B18, respectively. Red circles are measurements for the XSG stack (see the text). Observed COs are 
    compared to predictions of E-MILES SPS models: solid (dotted) pink and purple lines are models with solar (super-solar) metallicity and bimodal IMF of slopes $1.3$ and $3.0$, corresponding to a MW-like and bottom-heavy IMF, respectively. The (small) green arrows indicate the effect of adding a fraction of an intermediate-age E-MILES SSP on top of an old SSP (see the text); the blue arrow is the same but with an enhanced contribution of AGB stars for the intermediate-age component. The black arrow is the equivalent of the green arrow for the M05 models. The solid and dotted khaki (orange) arrows mark the effect of the ``empirical corrections'' to E-MILES SSPs with solar and super-solar metallicities, respectively,  for a MW-like (bottom-heavy) IMF. The brown and violet arrows show the effect of \aFe\ and \CFe\ abundance ratios on CO indices. The indices are measured after smoothing all data and models to a velocity dispersion of $360$~\kms. }
    \label{fig:fig4}
\end{figure*}

\begin{figure*}
\centering
	\includegraphics[width=\linewidth]{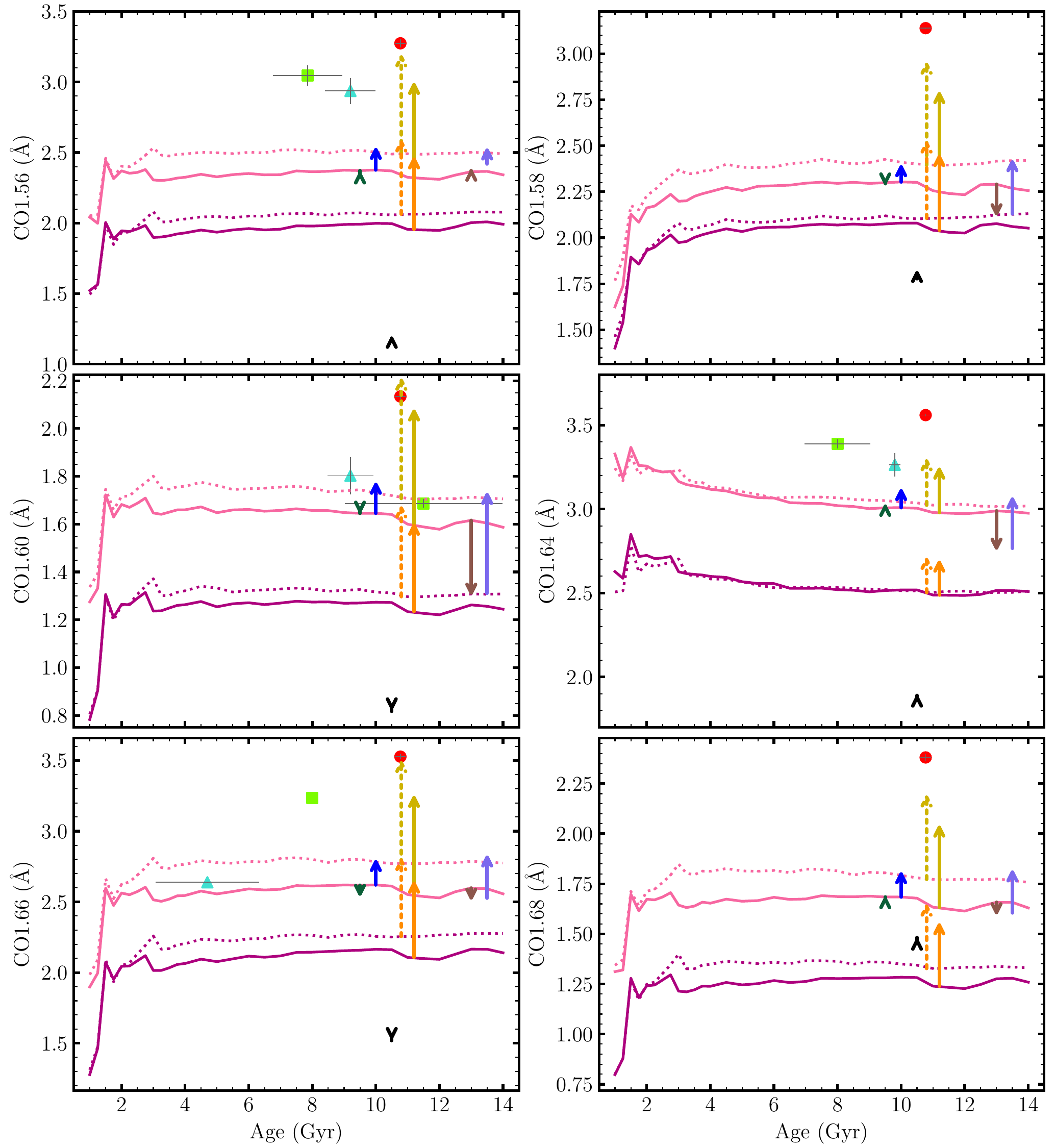}
    \caption{Same as Fig.~\ref{fig:fig4} but for \textit{H}-band indices}
    \label{fig:fig5}
\end{figure*}

To gain insight into the origin of the discrepancy between observed NIR CO features and model predictions, we scrutinise the effect of varying abundance ratios in the models (Sec.~\ref{sec:abundances}), as well as that of an enhanced contribution from AGB stars (Sec.~\ref{sec:AGB}).
The main results of this analysis are shown in Figs.~\ref{fig:fig4} and \ref{fig:fig5}, for \textit{K}- and \textit{H}-band indices, respectively. The figures are the same as Figs.~\ref{fig:fig2} and \ref{fig:fig3}, but showing only median values of line-strengths for different samples, as well as CO indices for the XSG stack. To avoid crowding the figure, only E-MILES models are plotted, with different arrows showing the effect of varying different parameters in the models, as detailed below. Note, also, that in Fig.~\ref{fig:fig4}, we do not  include the panel for D$_{\rm CO}$ (as in Fig.~\ref{fig:fig2}), as it does not add any further information with respect to the CO2.30 index, whose Lick-style definition is more similar to that of the other CO indices. 

\subsection{Abundance ratios}
\label{sec:abundances}

So far, we have considered only models constructed with stars following the chemical pattern of the solar neighbourhood. However, differences in the depth of CO features between models and data might also arise because of variations in \aFe, or other elemental abundances, with respect to a scaled-solar composition. According to \citealt{eftekhari2021} (see column "d" of their figures~4 and~6), the maximum change in the strength of CO indices due to elemental abundance variations comes from carbon and $\alpha$-enhancements. Therefore, in Figs.~\ref{fig:fig4} and \ref{fig:fig5}, we show the effect of $\alpha$-enhancement, based on $\alpha$-enhanced E-MILES models with an age of $13$~Gyr (brown arrows), and that of an enhancement in carbon abundance, based on CvD12 models with an age of $13.5$~Gyr and a Chabrier IMF (see the violet arrows). The violet and brown arrows correspond to variations of $\delta$\CFe\ = $+0.15$~dex and $\delta$\aFe\ = $+0.4$, respectively, which can be representative of massive ETGs, such as those of the XSG sample (see, e.g. \citealt{labarbera2017}). Note that we also have used the updated version of CvD12 models to see the effect of C-enhancement. Since the size of the violet arrow did not change with respect to that of CvD12, we do not include it in the plots to avoid crowding the figure. Moreover, since we are interested in the relative response of models to variations of \CFe\ and \aFe, we shifted the starting point of the violet arrow to the end point of the brown arrow.

As expected, increasing \CFe\ results in stronger CO lines in all cases, pointing in the right direction to match the data. However, the effect is counteracted by that of $\alpha$-enhancement (except for CO1.56, where the effect of $\alpha$-enhancement is negligible), so that, for all the \textit{K}-band CO indices,  the brown and violet arrows tend to cancel each other for the adopted abundance variations in the CvD12 models. Hence, the picture emerging from Fig.~\ref{fig:fig4} is similar to that of Fig.~\ref{fig:fig2}. For CO2.30 and CO2.35, even considering the effect of varying abundance ratios, the models are not able to match the observations, especially in the case of a bottom-heavy IMF. For CO2.32, the median index values for the F19 and B18 samples can be matched, but only with models having a MW-like IMF. 

For the \textit{H}-band CO line-strengths (Fig.~\ref{fig:fig5}), carbon abundance has a larger effect than \aFe, compared to \textit{K}-band indices, i.e. the relative size of violet vs. brown arrows in Fig.~\ref{fig:fig5} is larger than in Fig.~\ref{fig:fig4}. This shows, once again, the importance of studying lines from the same chemical species (CO) at different wavelengths (\textit{H} and \textit{K}). Even so, summing up the violet and brown arrows in Fig.~\ref{fig:fig5} does not allow us to reach the high CO values of massive XSGs. For instance, in the case of CO1.58 and CO1.60, summing up the effect of \CFe\ and \aFe\ would result in a (modest) increase of $\delta$(CO1.58)~$\sim0.1$~\AA\ and $\delta$(CO1.60)~$\sim0.2$~\AA, respectively, these variations being far smaller than the deviations between MW-like IMF E-MILES models and the XSGs stack, corresponding to $\sim0.9$~\AA\ and $\sim0.5$~\AA\ for CO1.58 and CO1.60, respectively. Note also that indices, such as CO1.56, for which the effect of abundance ratios is smaller (compared to, e.g., the effect of varying the IMF) do not show smaller deviations of data compared to models. In other terms, also a qualitative comparison of data and model predictions, seems to point against the effect of abundance ratios as the main culprit of the CO mismatch problem. However, one should bear in mind that the effect of abundance ratios on SSP models relies completely on theoretical stellar spectra, as well as molecular/atomic line lists, that are notoriously affected by a number of uncertainties, particularly in scarcely explored spectral regions, such as the NIR. Hence, we cannot exclude that the effect of \aFe\ and \CFe\ is underestatimated by current SPS models. Alternatively, one should seek for other possible explanations, as discussed in the following section.

\subsection{Intermediate-age stellar populations}\label{sec:AGB}

Since stars in AGB phase contribute most to the \textit{K}-band luminosity of stellar populations with ages between 0.3 and 2~Gyr ($\sim$30\% in E-MILES models, with \citet{maraston2005} models having the largest contribution from AGB stars among other models, i.e. $>$70\%), it has been suggested that the deep CO band-heads of ETGs in the \textit{K} band are due to the presence of AGB stars, from intermediate-age stellar populations \citep[e.g.][]{mobasher1996, mobasher2000, james1999, davidge2008, marmol2009}. It is important to assess, in a quantitative manner, if this hypothesis can account for the observed high CO line-strength values. To this effect, we first fit the observed XSG stacked spectrum with different SSP models, assuming a non-parametric star formation history (SFH), and then simulate the effect of intermediate-age populations by constructing ad-hoc two-component models.

To fit the XSG stack, we use the software {\sc STARLIGHT} \citep{cid2005}, a full spectral fitting code that allows us to fit a galaxy spectrum with a generic linear combination of input model spectra, i.e. the so-called ``base'' spectra. First, we use scaled-solar E-MILES SSPs as a base, including models with different ages and metallicities, and a Kroupa-like IMF~\footnote{Including bottom-heavy SSPs does not improve the {\sc STARLIGHT} fits significantly, as expected by the fact that CO line-strengths get weaker for $\rm \Gamma_b=3$ (see Sec.~\ref{sec:spectral_indices}).}. Note that this approach does not make any assumption about the SFH, which is treated in a non-parametric way. Hence, the effect of young populations is taken into account in the most general manner, without any restriction from the optical range. The {\sc STARLIGHT} fitting was carried out in the \textit{H} band, as CO absorptions dominate this spectral range. Figure~\ref{fig:fig6} compares the stacked spectrum of XSGs (black line), with the best-fitting composite stellar population model of E-MILES SSPs (pink line). The best-fitting model shows deviations at a level of $\sim2$\% from the stacked spectrum in the region of the CO bandpasses. Note that this is similar to what was found when comparing individual E-MILES SSPs to the XSGs' stack (see the fiducial model, plotted as a pink line, in Fig.~\ref{fig:fig1}). Since in the {\sc STARLIGHT} fits, there is no constraint on the age of the best-fitting SSPs, these results show that young populations do not help to resolve the tension between observations of CO lines and model predictions.

Based on near-ultraviolet (NUV) photometric data, \citet{yi2005} found that roughly $15$\% of bright ETGs at z $<0.13$ show signs of young  ($\lesssim1$~Gyr) populations at the level of $1$\%--$2$\% mass fractions. Also, \citet{schiavon2007} generated two-component models, showing that a mass fraction of the young component of $\sim0.5$\%--$1$\% provides a reasonably good match to the blue indices of nearby ETGs. This result has been recently confirmed, based on a combination of NUV and optical absorption lines for the XSGs, by \citet{salvador2021}, who found that the centre of massive ETGs are populated by a $0.7$\% mass fraction of stars formed within the last $1$~Gyr. 

To test this scenario, we contaminated the light of an old ($10$~Gyr) E-MILES SSP by a small fraction ($3$\% in mass) of an intermediate-age ($1.5$~Gyr) E-MILES SSP. The effect is shown for a solar metallicity and MW-like IMF population by the green arrows at $\sim10$~Gyr in Figs.~\ref{fig:fig4} and~\ref{fig:fig5}. Indeed, the arrows are small for all CO indices and, for CO1.58, CO1.60, and CO1.66,  they also point in the ``wrong'' direction, i.e. that of decreasing (rather than increasing) model line-strengths. However, when considering an E-MILES model with age of $1.5$~Gyr, solar metallicity, and a MW-like IMF, AGB stars contribute by $\sim1/3$ to its bolometric luminosity, while such fraction is larger for M05 models. Hence, one might attribute the small effect of the intermediate-age population to the less emphasized contribution of AGB stars to E-MILES, compared to M05, young SSP models. To address this issue, we used the AGB-enhanced version of E-MILES SSPs constructed by \citet{rock2017}. They calculated an AGB-enhanced E-MILES model of $1.5$~Gyr, solar metallicity and Kroupa-like IMF by using ``partial SSPs'', i.e. computing two SSPs, one by integrating stars along the isochrone without the AGB phase, and the other one by integrating only AGB stars, and combining the two models by assigning $70$\% luminosity-weight to the model made up of AGB stars only. This synthesised AGB-enhanced stellar population is added on top of an old population of $10$~Gyr assuming a $3\%$ mass fraction. The effect on CO line-strengths is shown by the blue arrows at $10$~Gyr in Figs~\ref{fig:fig4} and \ref{fig:fig5}. Although the blue arrows are larger than the green ones, they are not large enough to fit the median values for the F19 and B18 samples, except for CO2.35, CO1.60, and CO1.66. In the case of CO2.35 (see Fig~\ref{fig:fig4}), the median value of the F19 sample could be matched with a metal-rich and MW-like IMF E-MILES model, if one assumes that the effects of the blue, violet, and brown arrows (i.e. AGBs + \CFe\ + \aFe) sum up to $\sim0$. For CO1.60 and CO1.66, the comparison is limited by the small number of galaxies available for the F19 and B18 samples. As a further test, we added a $3$\% mass fraction of a $1.5$~Gyr M05 SSP with solar metallicity and Kroupa-like IMF to an old ($10$ Gyr) M05 SSP. The effect is shown by the black arrows at $\sim10$~Gyr in Figs~\ref{fig:fig4} and \ref{fig:fig5}. The effect of adding an emphasized-AGB intermediate-age population on top of an old one turns out to be negligible, and for the CO1.60 and CO1.66 indices, it goes also into the opposite direction compared to the data. Note also that there is no way the AGB-enhanced models can consistently match the strong CO line-strengths of the XSGs, in the \textit{H} and \textit{K} band.

As a final remark, we emphasize that our analysis does not rule out the presence of intermediate-age populations in ETGs, but it points against a scenario where the observed strong CO absorptions are mainly due to the presence of intermediate-age populations (i.e. AGBs) in these galaxies.

\section{An empirical modelling approach}\label{sec:empirical_approach}

\subsection{Searching for stars that match the strong CO lines}\label{sec:fitting}

\begin{figure*}
\centering
	\includegraphics[width=\linewidth]{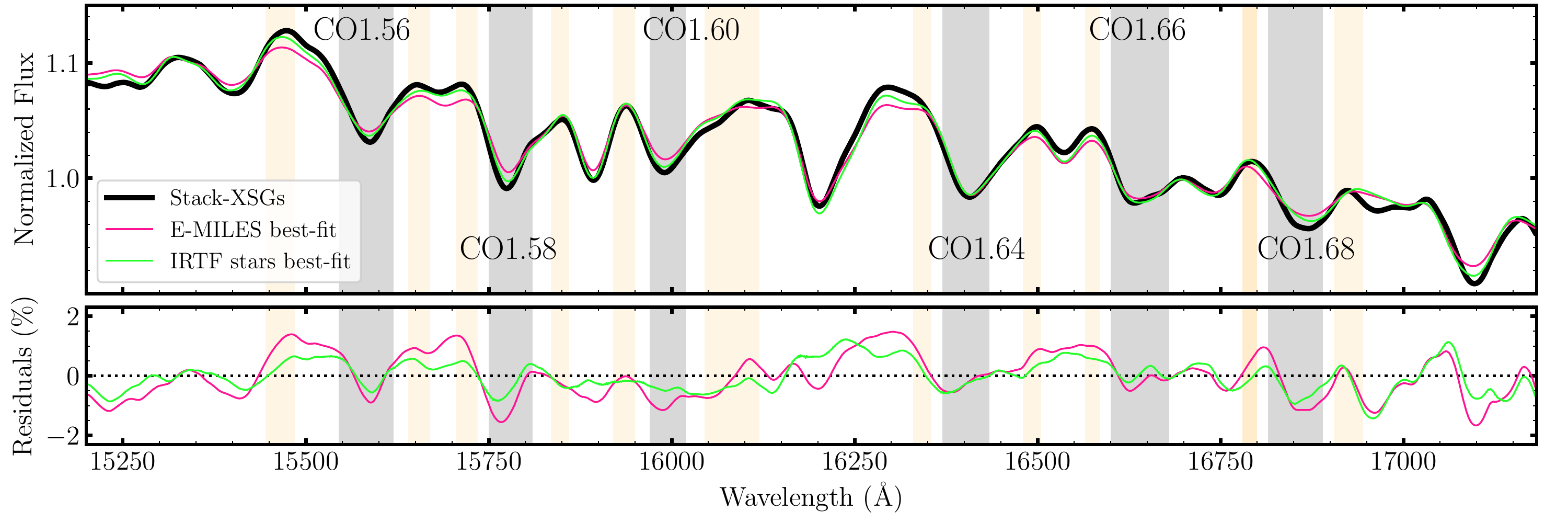}
    \caption{The upper panel shows the stacked spectrum of XSGs (black) in \textit{H} band, together with the best-fitting spectra obtained with E-MILES SSPs (pink) and IRTF stars (lime), running the spectral fitting code {\sc STARLIGHT} (see the text). CO features, with central absorptions and their pseudo-continua, are plotted as grey and orange shaded areas. All spectra have been normalised to their median flux. The lower panel shows the relative residuals of the stack with respect to each best-fitting spectrum. Notice the improvement in the matching of CO features when using IRTF stars, despite to the fact that only four stars received a non-negligible weight in the {\sc STARLIGHT} best-fit mixture (see the text).}
    \label{fig:fig6}
\end{figure*}

In order to identify the stars that might be responsible for the strong CO absorption observed in massive ETGs, we fitted the stacked spectrum of XSGs with {\sc STARLIGHT} (see above), using as an input base all the 180 individual stars of the IRTF library that are used to construct E-MILES models in the NIR. As for the fitting with E-MILES SSPs (see Sec.~\ref{sec:AGB}), we fitted only the \textit{H}-band region, where the signal from CO features is prominent with respect to that of other absorptions. The best-fit mixture of IRTF stars is shown in Fig.~\ref{fig:fig6}, as a lime-colored curve. The relative residuals between observed and model spectrum are shown in the bottom panel of the same figure. By comparing the residuals for the best-fit of IRTF stars with that for E-MILES SSPs (pink curve), it can be seen that using the stars improves significantly the fit to the observed spectrum, with residuals in the CO lines at the level of $\sim1$\%, i.e. about half of those for the E-MILES best-fitting model. Although some improvement in the fitting may be actually expected when employing the IRTF stars, given that this band is populated with so many CO absorptions, it is still remarkable how much smaller the obtained residuals are. Indeed, no improvement at all would be achieved in the case where the input stellar library completely lacked those stars responsible for CO absorption. {\sc STARLIGHT} also returns the weight (in light) of each star in the best-fit mixture. Surprisingly, we found that only $4$ (out of $180$) stars received a significant weight (> $0.5\%$) in the best-fit spectrum, namely HD~219734, HD~10465, HD~36003, and HD~187238. The light-weighted contribution of these stars from {\sc STARLIGHT} and their main stellar parameters from \citet{rock2016} are summarized in Tab.~\ref{tab:tab1}. HD~36003 is a dwarf star (low-mass star), while the other three stars are giants (massive stars). Hereafter, we refer to these stars as the \textit{H}-band best-fitting stars.

\strutlongstacks{T}
\begin{table}
	\centering
	\caption{ Properties of four IRTF stars that best-fit the XSG stacked spectrum in \textit{H} band.  Columns~1 and~2 give the name of each star and its weight in light to the best-fit {\sc STARLIGHT} model (see the text).  The effective temperature, surface gravity, and  metallicity of the stars are given in Columns~3, 4, and 5, respectively.  }
	\label{tab:tab1}
	\begin{tabular}{lcccr} 
		\hline\hline
		
		Star & Weight & \Teff & \logg	& \Feh \\
		     & ($\%$) & ({\small K})	         & (dex)	& (dex) \\
		(1) & (2)	& (3)	& (4)	& (5) \\
	    \hline
	    
HD 219734 & 43  & 3730 & 0.9 & 0.27  \\

HD 36003 & 18 & 4465 & 4.61 & 0.09\\

HD 187238 & 17 & 4487 & 0.8 & 0.177 \\

HD 10465 & 14 & 3781 & 0.5 & -0.458  \\
        
		\hline
	\end{tabular}
\end{table}

\begin{figure*}
\centering
	\includegraphics[width=0.87\linewidth]{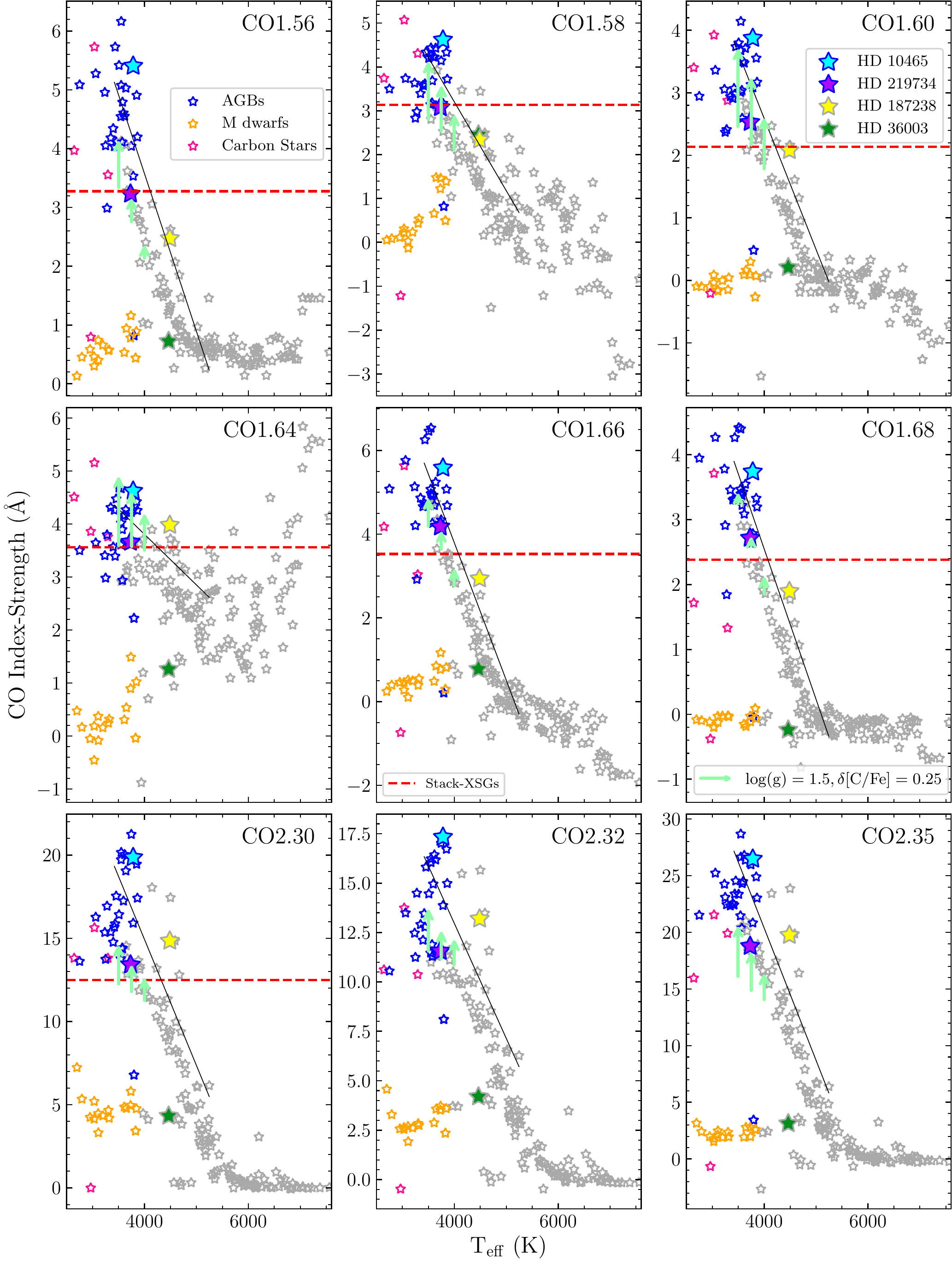}
    \caption{Index strength of CO indices for IRTF stars as a function of effective temperature. Indices have been measured on spectra smoothed at the common resolution of $\sigma=360$~\kms. Blue, orange, and pink colours correspond to AGB, M-dwarf and carbon stars at \Teff\ $< 3900$~{\small K}, respectively. Remaining stars are shown with grey colour. The four stars that contribute most to the net flux of the {\sc STARLIGHT} best-fitting model (see the text) are marked with filled star symbols of different colours (see labels in the top--right panel). Black lines are the linear fit to the CO-strong  sequence of giant stars, with \Teff $< 5500$~{\small K} (see the text for details). In each panel, the CO line-strength for the XSG stack is marked with a dashed-red line. The light green arrows mark the increase caused by a \CFe\ enhancement of $0.25$~dex, from the theoretical stars of \citet{knowles19_Thesis}, all having solar metallicity and \logg\ = $1.5$.}
    \label{fig:fig7}
\end{figure*}

Since the XSG stack is best-fit by only four stars, these stars have to be ``special'' somehow, and their properties might help up to shed light on the nature of the CO absorptions. To address this point, we measured the line-strength of CO indices for the spectra of all IRTF stars, and marked the position of the \textit{H}-band best-fitting stars in the CO vs. effective temperature  (\Teff) plots in Fig.~\ref{fig:fig7}. In this figure, different colours show different types of stars, according to the classification provided in table~2 of \citet{rock2015}, with blue and orange colours corresponding to AGB and M-dwarf stars, respectively. The five carbon stars in the IRTF library are shown in pink colour. Note that this classification is only available for stars cooler than $3900$~{\small K}. The remaining IRTF stars are plotted with grey colour. Figure~\ref{fig:fig7} suggests that those stars with \Teff\ $< 5500$~{\small K}, that are not classified as carbon stars and M-dwarfs, seem to split into two sequences. Most of the stars trace a well-defined, narrow sequence, that we call  the ``normal'' CO sequence throughout the paper, where the star HD~219734 (one of the \textit{H}-band best-fitting stars, see above) can be actually found, for {\it all} CO plots. Along this sequence, the CO line-strengths increase with decreasing \Teff. However, some  stars do not fall onto this sequence, but form a sort of  ``CO-strong'' sequence (where two of the \textit{H}-band best-fitting stars, namely HD~10465 and HD~187238, can be found in {\it all} CO plots)\footnote{Note that for CO1.58 and CO1.64, a double-branch sequence is not so clear. However, this might result from some sky residuals in the wavelength range of the CO lines, or some contamination of the CO lines from different absorbers. According to panel (a) in figure~A2 of \citet{eftekhari2021}, the central bandpass of CO1.58 is severely contaminated by telluric absorption lines and its red bandpass is contaminated by a strong emission line at $\sim15844$~\AA. Moreover, a magnesium line contributes to this absorption feature. In the same figure, in panel (b), the presence of two strong emission lines can be seen in both blue and red bandpasses of the CO1.64 index. The central feature also has some contribution from atomic silicon lines.}. To guide the eye, we performed a linear fit to the CO-strong sequence (see App.~\ref{sec:appendixA} for details), and marked such a sequence with black segments in Fig.~\ref{fig:fig7}. In all panels, we show the CO line-strengths for the XSG stack as horizontal red-dashed lines. These lines intersect the locus of stars at an effective temperature of about $4000$~{\small K}. This is the temperature where stars in the CO-strong sequence deviate the most from those in the normal sequence.

We also attempted to fit the spectrum of the \textit{H}-band best-fitting stars with the MARCS \citep{gustafsson2008} library of very cool stellar spectra, and tried to extract additional information regarding their $\alpha$ abundance ratio. However, the best fitting models do a poor job of predicting the shape of the spectra of such cool stars and CO indices; therefore the derived parameters are less reliable. Here, we only mention that the results point to a lower $\alpha$ abundance for stars in the CO-strong sequence compared to the one in the normal sequence (see App.~\ref{sec:appendixB} for details of this experiment).

As noted above, two stars that were assigned the highest weight in the {\sc STARLIGHT} best-fitting spectrum (HD~10465 and HD~187238; see Tab.~\ref{tab:tab1}) occupy the CO-strong sequence, while the other two (HD~219734 and HD~36003) follow the normal sequence. This suggests that in order to match the observed spectrum of ETGs, a significant contribution from the CO-strong sequence might be required. However, as a general caveat, one should notice that massive ETGs might contain different stars than the few CO-strong-sequence stars in the IRTF library. An interesting point is that the two stars in the higher sequence have almost the same \Teff\ as the two other stars. This may explain why SPS models do actually fail to reproduce CO features. A nominal SSP model averages the available stellar spectra along the isochrones. Since stars in the normal sequence are in larger number compared to those in the CO-strong sequence, the contribution from the latter is diluted in the synthesised models. The {\sc STARLIGHT} fitting results are suggesting, instead, that it should be the other way around, with a large weight from the CO-strong sequence. In order to remedy this situation, we constructed ad-hoc SPS models, as detailed below.

\subsection{Empirical corrections to E-MILES models}\label{sec:empirical_model}

\begin{figure*}
\centering
	\includegraphics[width=\linewidth]{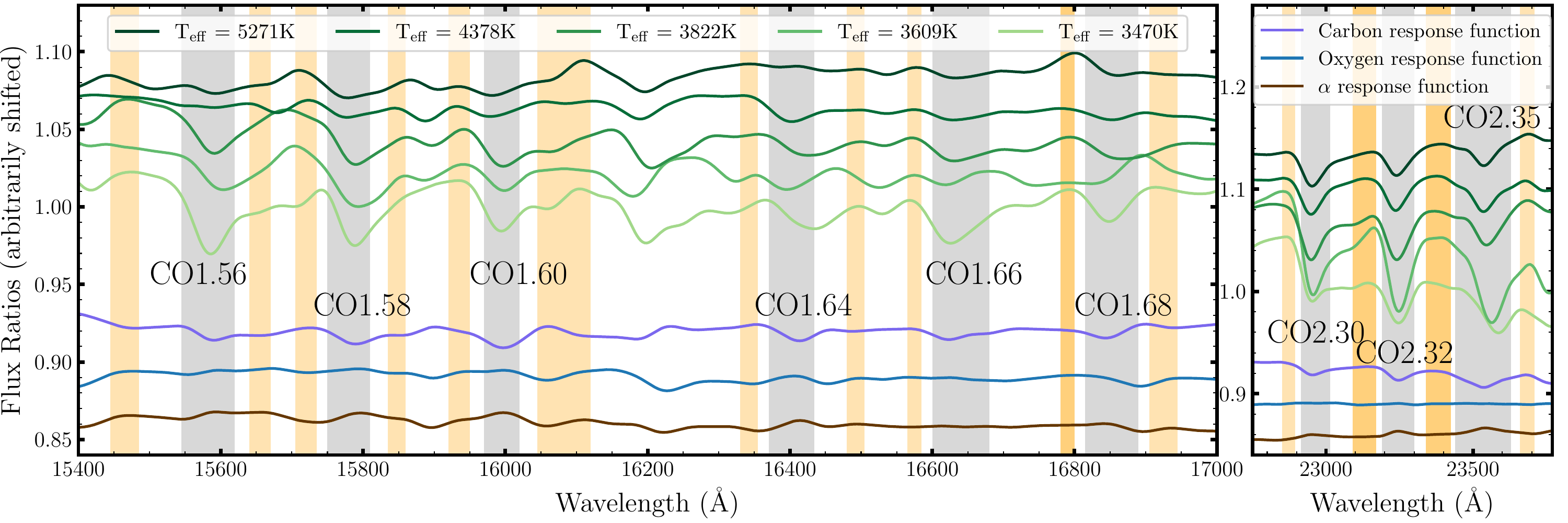}
    \caption{Comparison of empirical CO responses (light through dark green curves) with \CFe\ (violet), \OFe\ (blue), and \aFe\ (brown) responses from CvD12 models. The light to dark green spectra are defined as the ratio between mean spectra of stars in the CO-strong and normal CO sequences, for stars with different temperature. The definition of indices are overplotted with grey (central bandpasses) and orange (pseudo-continua bandpasses) areas.
    The left and right panels refer to \textit{H} and \textit{K} band, respectively. All spectra have been smoothed to $\sigma=360$~\kms. }
    \label{fig:fig8}
\end{figure*}

We modified E-MILES stellar population models by shifting stars in the normal CO sequence to those in the upper (CO-strong) one. The procedure is described in detail in App.~\ref{sec:appendixA}. In short, we systematically separated giant stars into the two sequences (according to all the available CO indices), and for stars that share similar stellar parameters, we divided the mean spectrum for the CO-strong sequence with that for the normal sequence, to obtain a (multiplicative) differential response of ``CO-enhancement'', as illustrated in Fig.~\ref{fig:fig8} (see light through dark green spectra). We point out that this procedure is possible because a number of stars are in the upper sequence for all the CO indices, i.e. there is actually a population of stars, with strong lines, that can be singled out from the normal sequence, which extend over a range of temperatures. The responses obtained in this way were interpolated at different temperatures, and applied to the giant star spectrum in the normal sequence. New SSP models were synthesised accordingly, using the ``empirically corrected'' giant stars, for an age of $11$~Gyr, \Mh\ = $+0.06$ and $+0.26$, and $\Gamma_{b} = 1.3$ and $3.0$, respectively, over a wavelength range from $15400$ to $23800$~\AA. 

We measured the CO line-strengths on the empirically corrected models. In Figs.~\ref{fig:fig4} and \ref{fig:fig5}, we show the variation of CO indices, compared to the reference E-MILES models, as khaki and orange arrows, for IMF slopes of $\Gamma_{b} = 1.3$ and $3.0$, respectively. Solid and dotted arrows correspond to solar and super-solar metallicity models, respectively. The empirically-corrected SSPs have significantly larger CO indices. In the case of CO2.30, CO1.60, and CO1.66, the khaki arrows would allow one to fit the stacked spectrum of XSGs. However, since the IMF has been shown to be bottom-heavy for these galaxies, it should be looked at the orange arrows, whose size is not enough to match the data.  For some \textit{H}-band indices, i.e. CO1.56, CO1.60, and CO1.66, the khaki arrows predict even larger CO values that the median line-strengths for the samples of B18 and F19. In the case of CO1.58, CO1.64, and CO1.68, although the empirically corrected models improve the predictions of CO indices, they cannot match those of massive ETGs (even for a MW-like IMF).

We note that the dotted arrows have approximately the same size as solid arrows, i.e. the effect of the empirical correction does not depend on metallicity. Perhaps, this is not surprising, as stars of the IRTF stellar library are biased towards solar metallicities~(see \citet{rock2017} and references therein).  Also, khaki and orange arrows have approximately similar size, implying that the empirical CO response is not coupled to that of IMF, as it is the case for Na-enhanced E-MILES models (see~\citealt{labarbera2017} for details).  This stems from the fact that the CO correction is only performed on giant stars, while a bottom-heavy IMF increases the number of dwarf, relative to giant, stars. 

\subsection{What is driving the empirical corrections?}\label{sec:drivers}

The empirically corrected models provide closer predictions to observed CO indices compared to E-MILES models, although still far from a perfect match. In order to make further progress, we tried to understand the physical drivers behind the empirical corrections, to possibly tune the models further. 

Figure~\ref{fig:fig8} shows the CO-response functions of five selected stars, with different \Teff\ but otherwise similar stellar parameters, that we used to construct the empirically corrected models (see App.~\ref{sec:appendixA} for details). The violet, blue, and brown lines in the figure plot responses corresponding to a carbon enhancement of $0.15$~dex, an oxygen enhancement of $0.2$~dex, and an $\alpha$ enhancement of $0.2$~dex, based on CvD12 SSP models (with an age of $13.5$~Gyr, \Feh\ = $0.0$, and Chabrier IMF), respectively. Indeed, the empirical responses look more similar to those for carbon enhancement, although the differences in the depth of CO indices are quite significant. On the contrary, the response to oxygen enhancement is very different from the empirical responses, being almost flat in the regions of CO absorptions. The effect of $\alpha$ enhancement is even more dissimilar, as it shows bumps in the regions of the CO central passbands, consistent with the fact that CO line-strengths anti-correlate with \aFe\ (see Sec.~\ref{sec:abundances}). 

Therefore, we speculate that the empirical corrections might be reflecting the effect of carbon enhancement on (cool) giant stars.  To further test this hypothesis, the \CFe\ of stars in the CO-strong and normal sequences should be compared. Unfortunately, carbon abundances for giant stars in the IRTF library have not been measured yet. Hence, we relied on theoretical C-enhanced stars from \citet{knowles19_Thesis} to see if an enhancement in carbon abundance might explain the difference between CO line-strengths in the two CO sequences. In Fig.~\ref{fig:fig7}, we show the effect of a \CFe\ enhancement of $0.25$~dex on theoretical cool giant stars with light-green arrows. Interestingly, the size of the arrows increases with decreasing \Teff. However, one should bear in mind that theoretical stellar spectra are rather uncertain for very cool stars, and these models stop at $3500$~{\small K}. Indeed, we may expect that this trend continues at lower \Teff, with CO absorptions getting even stronger. Focusing only on the difference due to enhancement and neglecting the starting point, we see that the arrows are comparable to the CO line-strengths difference between normal and CO-strong sequence stars. For CO1.58, CO1.60 and CO2.32 indices, the arrows can bring the stars from the normal to the CO-strong sequence, while for other indices, the arrows can explain only part of the difference between the two sequences, with a larger gap for cooler stars. For instance, the arrow at 3500K for CO1.68 is too small, and it is unable to reach a group of three stars, with \Teff\ $\sim3500$~{\small K} and CO1.68 as high as $\sim4.4$~\AA.  

\begin{figure*}
\centering
	\includegraphics[width=0.89\linewidth]{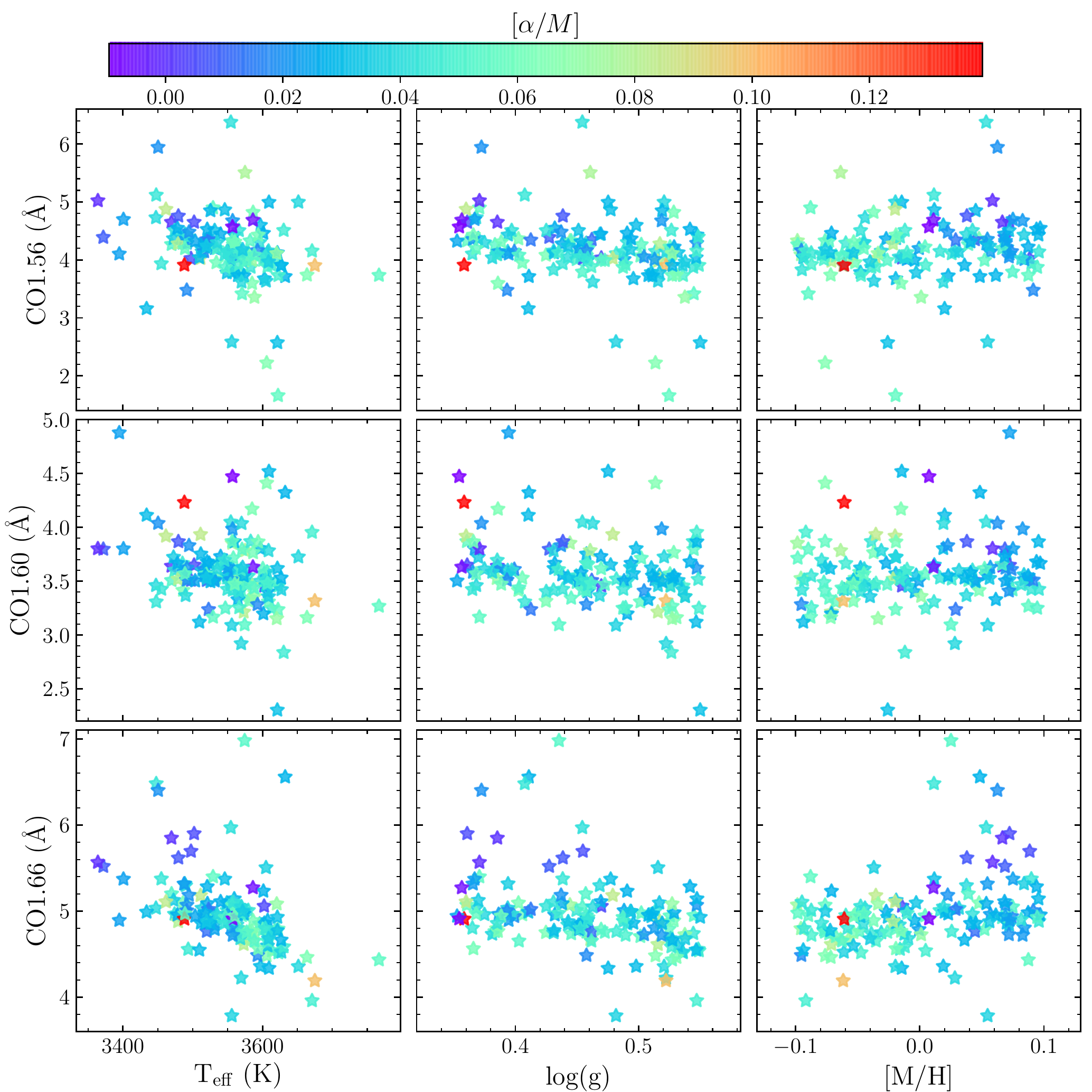}
    \caption{CO indices measured on the spectra of APOGEE stars as a function of stellar parameters, namely \Teff\ (left), \logg\ (middle), and \Mh\ (right). The stars are coloured according to \aFe, as shown from the colourbar on the top.  }
    \label{fig:fig9}
\end{figure*}

As shown in Figs.~\ref{fig:fig4} and~\ref{fig:fig5}, increasing \aFe\ abundance causes the CO absorptions to weaken. However this prediction, which is qualitatively similar in both E-MILES $\alpha$-enhanced and CvD12 models, is in contrast with predictions of A-LIST SSP models \citep{ashok2021}. According to their figure~6, CO absorptions get stronger by increasing \aFe. A-LIST provides fully empirical SSP model predictions, based on the APOGEE stellar library, while $\alpha$-enhanced E-MILES SSPs are  semi-empirical models (i.e. the relative effect of $\alpha$-enhancement is estimated through the aid of theoretical star spectra). Unfortunately, \aFe\ abundance ratios have not been measured for all IRTF stars. However, we further assessed the effect of \aFe\ on CO indices by looking at elemental abundance ratios for the APOGEE stellar library, as computed with ASPCAP. To this effect, we selected a set of APOGEE stars as described in Sec.~\ref{sec:libraries}. We attempted to single out the effect of surface gravity, metallicity, and carbon enhancement by only selecting stars within a narrow range of stellar parameters ($\rm 0.35 <$ \logg\ $< 0.55$, $-0.1 <$ \Mh\ $< 0.1$, and $\rm 0 <$ \CFe\ $< 0.05$). Since the wavelength coverage of APOGEE spectra is divided across three chips with relatively narrow ranges (blue chip from $1.51$ to $1.581~\mu$m, green chip from $1.585$ to $1.644~\mu$m, and red chip from $1.647$ to $1.7~\mu$m), we were able to measure line-strengths for only three CO indices (i.e. CO1.56, CO1.60, and CO1.66, respectively). Figure ~\ref{fig:fig9} shows CO line-strengths for APOGEE stars as a function of \Teff, \logg, and \Mh\ (left-, mid-, and right-panels), respectively, with stars being colour-coded according to their \aFe. According to the figure, stars with higher CO do not show a higher value of \aFe. In many cases, stars with high CO seem to have lower (rather than higher) \aFe. In the plot of CO1.60 vs. \Teff, only one star (in red) has high $\alpha$-enhancement and high CO1.60. For CO1.66, a correlation (anti-correlation) of the index with surface gravity (metallicity) is actually observed. Note that these results are in disagreement with predictions from the A-LIST models of \citet{ashok2021}, with the origin of such disagreement remaining unclear.

Overall, our analysis shows that it is very unlikely that $\alpha$ enhancement is the missing piece of the CO puzzle. On the other hand, the effect of carbon on low-temperature giant stars seems to be the most likely candidate to explain the strength of CO lines. However, the predictions from theoretical models should be improved, and extended to stars with lower temperatures ($\lesssim3500$~{\small K}) at high metallicity, in order to draw firm conclusions. 

\section{Discussion}\label{sec:discussion}

\begin{figure*}
\centering
	\includegraphics[width=16cm]{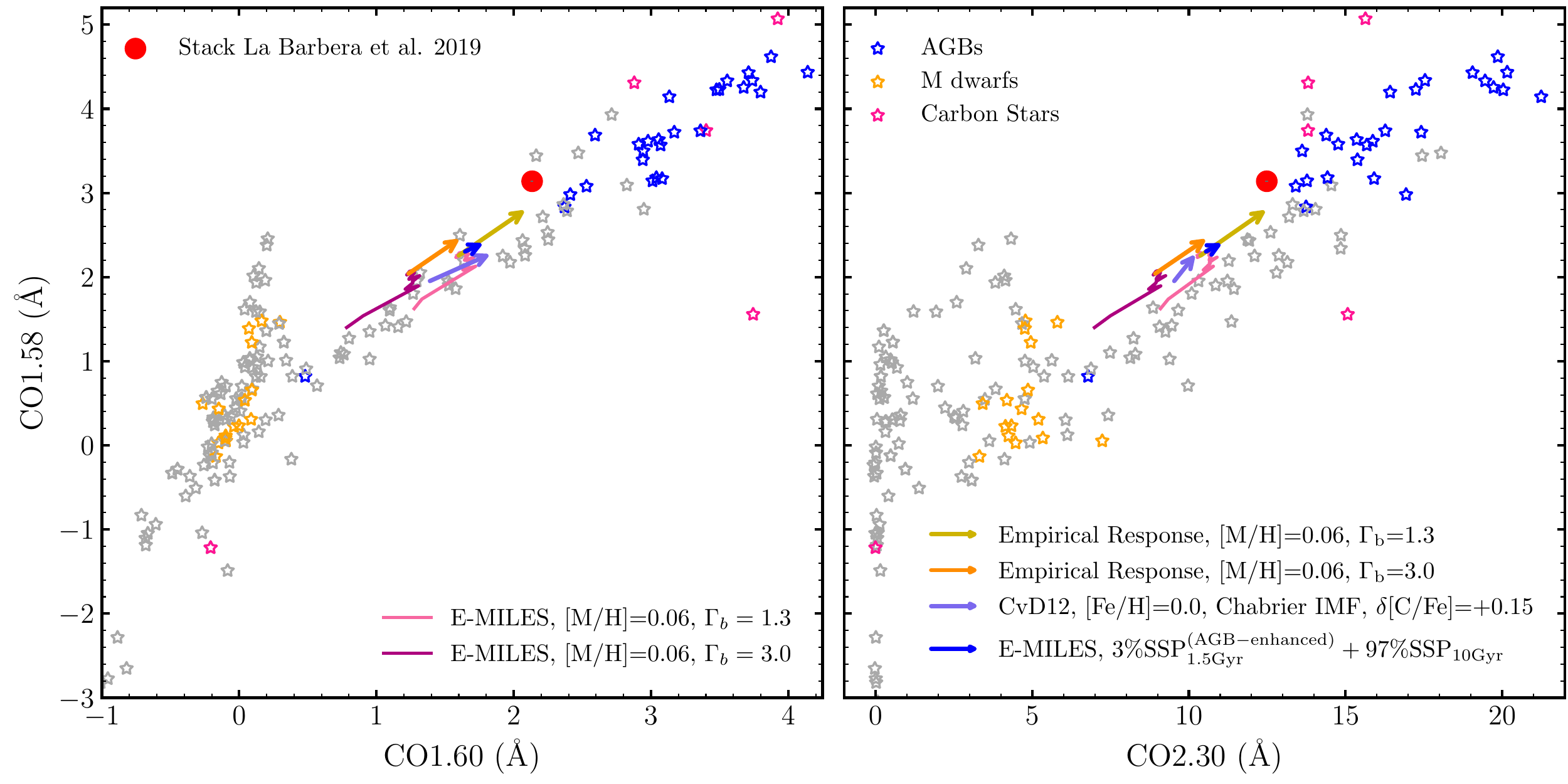}
    \caption{ A selected pair of CO vs. CO plots (CO1.58 vs. CO1.60 on the left; CO1.58 vs. CO2.30 on the right), showing IRTF stars (star symbols), the XSG stacked spectrum (filled red circle),  and predictions for solar-metallicity E-MILES SSP models having  ages from 1 to $14$~Gyr, for a MW-like and bottom-heavy IMF (see pink and purple lines), respectively. The violet arrows show the effect of increasing carbon abundance for CvD12 models, while the (small) blue arrows plot the increase of CO indices when accounting for AGB-enhanced intermediate-age population on top of an old stellar component. The effect of our empirical corrections on SSP model predictions is shown with khaki and orange arrows, for a MW-like and a bottom-heavy IMF, respectively. 
}
    \label{fig:fig10}
\end{figure*}

It is instructive to look at the mismatch between models and data using CO-CO diagrams, i.e. plotting one CO index against CO line-strengths for other features. In Fig.~\ref{fig:fig10}, we show two such diagrams, based on three CO indices (CO1.58, CO1.60, and CO2.30, respectively) as measured for IRTF stars (star symbols), the stacked spectrum of XSGs (red point), and E-MILES models with ages from $1$ to $14$~Gyr (see the pink and purple lines, corresponding to solar-metallicity models for MW-like and bottom-heavy IMF, respectively). In the CO vs. CO plots, the locus of stars is well defined, forming a relatively narrow strip. Interestingly, the point of massive galaxies falls off the main strip: for  values of CO1.60 $\sim2$~\AA\ and CO2.30 $\sim12$~\AA, stars have CO1.60 $\sim 2.4$~\AA, while the XSG stack has CO1.60 $\sim3.2$~\AA. As expected, E-MILES model predictions follow the locus of stars, predicting lower CO values compared to the data. However, to be able to match the XSG stack, the models should not only increase the CO line-strengths but, also, move away from the main strip of IRTF stars. 

Note that in both panels of Fig.~\ref{fig:fig10}, some dwarf stars (see the orange stars in the figure) do not share the same locus as the rest of the IRTF stars, but they are actually shifted to higher values of CO1.58 ($\sim1.4$\AA ), at given values of CO1.60 ($\sim0.2$\AA ) and CO2.30 ($\sim6$\AA ), respectively. The result is that predictions of models with a bottom-heavy IMF (orange arrows in the figure) are slightly above the main star locus. However, at the same time, these models predict lower values of CO ($\Delta$CO1.58 $\approx-0.2$~\AA, $\Delta$CO1.60 $\approx-0.5$~\AA, and $\Delta$CO2.30 $\approx-2$~\AA, comparing the tips of the orange and khaki arrows in the figure). Therefore, while bottom-heavy models worsen the gap between observed and model line-strengths, on the other hand, the CO-CO plots suggest that IMF variations might help to reconcile models and data by producing greater CO1.58 line-strengths than CO2.30.

In Fig.~\ref{fig:fig10}, we also show, with blue arrows, the effect of an AGB-enhanced population  (based on AGB-enhanced E-MILES SSPs; see Sec.~\ref{sec:AGB}), trying to mimic the presence of an intermediate-age population. The AGB-enhanced models increase CO1.58 by only $0.1$~\AA, while the discrepancy between the red point (i.e. the XSG stack) and the pink line (fiducial E-MILES model) is about $8$ times larger. Moreover, the change in CO1.60 (CO2.30) line-strength due to the blue arrow is $\sim0.1$ ($\sim0.5$)~\AA, i.e. one-fifth (one-forth) of the offset between the models and data. We conclude, as already discussed in Sec.~\ref{sec:AGB}), that while AGB stars might have a relevant contribution to the NIR light of (massive) galaxies, they are likely not responsible for the strong CO line-strengths in the \textit{H} and \textit{K} bands.

Khaki and orange arrows in Fig.~\ref{fig:fig10} are the same as in Figs.~\ref{fig:fig4} and~\ref{fig:fig5}, plotting the increase of CO line-strengths caused by the empirical correction on giant stars, for both MW-like and bottom-heavy IMF models, respectively. Both arrows increase the CO1.58, CO1.60, and CO2.30 line-strengths by $\sim0.4$, $0.4$, and $1.5$~\AA, respectively. While the khaki arrow (compared to the orange one) brings the model indices closer to the XSG stack, the orange arrow is not able to reach the data, but it is slightly off the star sequence, similar to the XSG stack. This suggests that in order to match the CO line-strengths, one would need an effect similar to that of the empirical corrections, plus the slight offset due to a bottom-heavy IMF. 

The effect of carbon enhancement from CvD12 models is also shown in Fig.~\ref{fig:fig10} (violet arrows), to be compared with the empirical responses. In the CO1.58 vs. CO1.60 diagram, the violet arrow, although increasing the CO indices (by $\sim0.4$~\AA\ for CO1.58, and $\sim0.4$~\AA\ for CO1.60), it is along the stellar locus, while in the CO1.58 vs. CO2.30 plot, the arrow seems to point to the correct direction to match the XSG stack (though with a small overall variation of only $\Delta$CO2.30 $\approx0.5$\AA). 

Similar to the effect of carbon-enhancement, it seems that carbon stars (pink stars in Fig.~\ref{fig:fig10}) might also be able to bring the models out of the stellar locus in the CO1.58 vs CO2.30 diagram, while this is not the case for the CO1.58--CO1.60 plot, as in the latter case, pink stars are somehow aligned to the sequence of blue stars. However, according to Fig.~\ref{fig:fig7}, for most CO vs. \Teff\ panels, carbon stars are not in the CO-strong sequence, i.e. they would not help in matching all the \textit{H}- and \textit{K}-band CO line-strengths. Again, this shows the importance of combining the largest available set of CO features, as we do in our analysis, and gives further support to our conclusion that adding an intermediate-age stellar population (that would also include carbon stars) to an underlying old component does not solve the issue with NIR CO spectral features.

Using optical and \textit{J}- and \textit{K}-band absorption features, including the first two CO bandheads in \textit{K} band, \citet{alton2017, alton2018} studied stellar population gradients in the spectra of eight massive ETGs. They showed that models that do not account for the effect of \CFe\ variations underpredict the CO bandheads in the \textit{K} band. Moreover, they showed that to fit H$_{\beta}$, in the optical, a large enhancement in carbon abundance is also required. In other terms, an over-abundance of carbon seems to have a prominent role in matching CO lines, in agreement with the suggestions from our analysis. However, \citet{alton2017, alton2018} also used CO features in the \textit{K} band to conclude in favour of a MW-like IMF in the center of (some) massive ETGs, in contrast to studies based on (optical) spectral features. Indeed, our analysis shows that current stellar population models in the NIR are still not accurate enough to allow for a quantitative matching of CO lines to be performed, and an even smaller effect to be constrained, such as that of a varying IMF. The results presented here demand a new generation of NIR stellar population models, after a significant effort to move beyond the current limitations of theoretical star spectra is made, particularly for the predictions of abundance ratio effects in low-temperature (giant) stars. Along the same lines, we point out that while our ad-hoc empirically-corrected SPS models do not match the observations yet, they tend to significantly reduce the discrepancy with respect to the observed CO strengths. Admittedly, the interpretation of our empirical corrections as an effect of \CFe\ for low-temperature giants remains rather speculative, but urges for their study and opens up new avenues for improving SPS models in the NIR spectral range.

\section{Summary and Conclusions}\label{sec:conclusions}

We have shown, in a comprehensive manner, that a whole set of CO lines in the spectra of massive ETGs, from \textit{H} through to \textit{K} band, lie significantly above the predictions of widely-used state-of-the-art SPS models. We have explored different possible reasons for this ``CO-mismatch'' problem, finding that an individual enhancement of carbon abundance might be the most likely explanation, compared to other scenarios (such as an enhanced contribution of AGB stars from intermediate-age populations). In general, our study highlights the importance of improving SPS models in the NIR, in the following aspects:

\begin{itemize}
    \item \textit{non-solar chemical abundances:}  we need substantial progress in the modelling of the response of stellar spectra to elemental abundance variations. In particular, the effect of varying abundance ratios for cool stars (\Teff\ $< 4000$~{\small K}) is far from being well understood, and might be crucial to explain the current issues with NIR spectral lines. In addition, since current SPS modelling for varying abundance ratios is based on scaled-solar isochrones,  we need a significant improvement on isochrones with non-solar abundances to create fully consistent SSP models with varying abundances. Moreover, the interplay between C, N, and O elements and their effects on CO indices are not yet fully understood. As some of the current models consider O as one of the $\alpha$-elements, a different treatment of O from $\alpha$-elements may be an interesting avenue for further investigation.   
    
   \item \textit{very cool stars:} current (theoretical) models struggle to reproduce atomic and molecular bands for stars with \Teff\ $< 3500$~{\small K}. Moreover, SPS would benefit from an improved treatment of evolved stellar evolution phases, such as those of red giants and supergiants, and the AGB, which have a prominent contribution to the NIR light of a stellar population in various age regimes.
   
    \item \textit{high-metallicity stars:} empirical stellar libraries, used to construct SPS models, are based on MW stars, having an unavoidable bias towards solar metallicity. Based on current SPS model predictions, CO indices do not depend significantly on metallicity, but from the study of individual stars and clusters \citep{aaronson1978, oliva1995}, CO indices are found to increase with increasing metallicity. Therefore, models with a good coverage of stars in the supersolar metallicity regime might yield further important clues to understand the CO-mismatch problem.
    
\end{itemize}

As a final remark, we would like to emphasize that a revision of SPS models in the directions suggested by the NIR CO indices, should take carefully into account the constraints provided by other spectral ranges, such as the optical and the UV. For example, fitting just a single CO line or a number of them, could lead to misleading derivations of stellar population properties.

\section*{Acknowledgements}

We are thankful to the reviewer, Dr. Russell Smith, for his careful reading of the manuscript and valuable comments. The authors acknowledge support from grant PID2019-107427GB-C32 from the Spanish Ministry of Science, Innovation and Universities (MCIU). This work has also been supported through the IAC project TRACES which is partially supported through the state budget and the regional budget of the Consejer\'\i a de Econom\'\i a, Industria, Comercio y Conocimiento of the Canary Islands Autonomous Community. 

\section*{Data Availability}

The E-MILES SSP models are publicly available at the MILES website (\url{http://miles.iac.es}). The updated version of Na-enhanced models of \citet{labarbera2017} are also available from the same website (under "Other predictions/data"). The \citet{conroy2012a} SSP models are available upon request to the authors (see \url{https://scholar.harvard.edu/cconroy/projects}). The \citet{conroy2018} models are available for download at \url{https://scholar.harvard.edu/cconroy/sps-models}. The \citet{maraston2005} are downloaded from  \url{http://www.icg.port.ac.uk/~maraston/Claudia's_Stellar_Population_Model.html}. Observations of \citet{labarbera2019} sample made with ESO Telescope at the Paranal Observatory under programmes ID 092.B-0378, 094.B-0747, 097.B-0229 (PI: FLB). The central spectra and stacked central spectrum are available from FLB upon request. The spectra of \citet{baldwin2018} sample were taken using the Gemini Near-Infrared Spectrograph on the Gemini North telescope in Hawaii through observing program GN-2012A-Q-22. The reduced spectra (FITS files) are available via \url{https://github.com/cbaldwin1/Reduced-GNIRS-Spectra}. Observations of \citet{francois2019} sample made with ESO Telescopes at the La Silla Paranal Observatory under programme ID 086.B-0900(A). The reduced spectra (FITS files) are available via \url{http://cdsarc.u-strasbg.fr/viz-bin/qcat?J/A+A/621/A60}. Observations of \citet{silva2008} sample performed at the European Southern Observatory, Cerro Paranal, Chile; ESO program 68.B-0674A and 70.B-0669A. Observations of \citet{marmol2008} sample performed at the European Southern Observatory, Cerro Paranal, Chile, as well. \citet{rock2017} corrected the data of these two samples to restframe and convolved them to a resolution of $\sigma=360$~\kms and they are available from EE upon request. The IRTF Spectral Library is observed with the SpeX spectrograph, at the NASA Infrared Telescope Facility on Mauna Kea and the spectra are publicly available at \url{http://irtfweb.ifa.hawaii.edu/~spex/IRTF_Spectral_Library/}. Theoretical stars are computed by ATK specifically for this project and are available from ATK upon request. APOGEE stars can be downloaded through the \url{https://www.sdss.org/dr16/} website.



\bibliographystyle{mnras}
\bibliography{references} 




\appendix

\section{Constructing empirically corrected models}\label{sec:appendixA}

\begin{figure*}
\centering
	\includegraphics[width=0.89\linewidth]{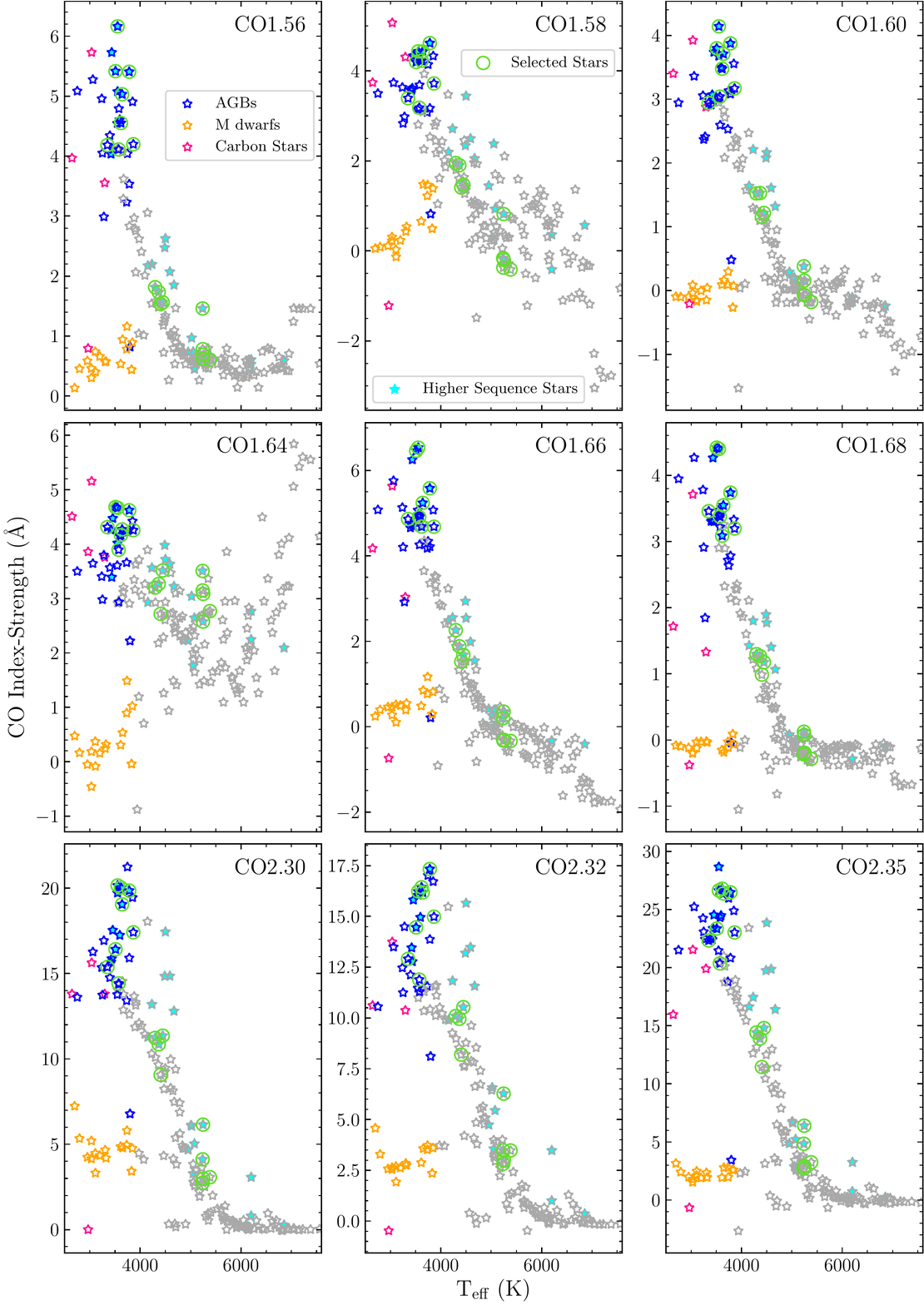}
    \caption{CO index-strengths of the IRTF library stars plotted as a function of temperature, all measurements performed at resolution of $\sigma=360$~\kms. Blue, orange and pink colours represent AGB, M-dwarf and carbon stars for \Teff\ $< 3900$~{\small K}, respectively. Stars with higher temperatures are shown in grey colour. Giant stars belonging to the CO-strong sequence are shown as filled star cyan symbols. The open lime circles show different sets of stars sharing similar parameters in the normal and CO-strong sequences, which were used to derive the empirical responses (see the text).}
    \label{fig:figA}
\end{figure*}

We explain here the procedure of identifying stars in the CO-strong sequence and the derivation of spectral response functions to bring giant stars in the normal sequence to this CO-strong sequence. 
We identified for each index the CO-strong giants through a sigma clipping procedure within temperature intervals of $200$~{\small K}. Those stars that were flagged in the CO-strong sequence for at least five CO indices were definitely classified as CO-strong, which are shown as filled cyan star symbols in Fig.~\ref{fig:figA}. Then we selected five sets of stars, sharing similar \Teff, \logg, and \Feh, within these two sequences, i.e. normal and CO-strong. These sets of stars are indicated with lime open circles in Fig.~\ref{fig:figA}, and their parameters are given in Tab.~\ref{tab:tabA}.

\strutlongstacks{T}
\begin{table}
	\centering
	\caption{Different sets of stars with similar stellar parameters selected for deriving empirical response functions as a function of temperature. Columns~1 and 2 indicate the group and sequence of stars to which they belong. Column~3 presents the ID of stars, and Cols.~4, 5, and 6 give the effective temperature, surface gravity, and metallicity.}
	\begin{adjustbox}{width=\columnwidth}
	\label{tab:tabA}
	\begin{tabular}{lccccr} 
		\hline
		
		Group & Sequence & Star & \Teff & \logg & \Feh \\
		      &          &   	& ({\small K})        	& (dex)  &	(dex) \\
		(1)   & (2)      & (3)	& (4)	        & (5)    &	(6)	  \\
	    \hline
\multirow{3}{*}{Group 1} & \multirow{2}{*}{CO-strong} & HD~14488  & 3509 & -0.16 & -0.77 \\
 &  & HD~14469   & 3551 & -0.16 & -0.69 \\
 
 & normal & HD~40239   & 3349  & 0.47  & 0.03\\

        \hline
\multirow{3}{*}{Group 2} & \multirow{2}{*}{CO-strong} & HD~35601  & 3617  & 0.06 & -0.20\\
 &  & HD~14404   & 3640  & 0.32 & -0.07\\
 
 & normal & HD~4408   & 3571  & 0.99 & -0.51\\
 
        \hline
\multirow{2}{*}{Group 3} & CO-strong & HD~10465   & 3781  & 0.50 & -0.46\\

 & normal & HD~23475   & 3863  & 0.78 & -0.63\\
 
        \hline
\multirow{4}{*}{Group 4} & \multirow{3}{*}{CO-strong} & HD~16068   & 4296  & 1.80 & -0.03\\
 &  & HD~164349   & 4446  & 1.50  & 0.39\\
 &  & HD~35620   & 4367  & 1.75 & -0.03\\
 
 & normal & HD~161664   & 4405  & 1.63 & -0.03\\
 
        \hline
\multirow{5}{*}{Group 5} & \multirow{2}{*}{CO-strong} & HD~179870   & 5246  & 1.96  & 0.11\\
 &  & HD~42454   & 5238  & 1.10  & 0.03\\
 
 & \multirow{3}{*}{normal} & HD~3421   & 5383  & 2.45 & -0.17\\
 &  & HD~74395   & 5250  & 1.30 & -0.05\\
 &  & HD~39949   & 5240  & 1.23 & -0.10\\

	\end{tabular}
	\end{adjustbox}
\end{table}

These sets of stars allowed us to compute a ``response function'' by dividing the mean spectrum of stars in the CO-strong sequence by the mean spectrum of stars in the normal sequence. These responses were interpolated to cover the temperature range of the giant stars in the normal sequence. Finally, we applied these response functions to all the giants in the normal sequence to correct them into the CO-strong sequence.

\section{Current Limitations of Stellar Models in the Low Temperature Regime}\label{sec:appendixB}

One way to get information on the \textit{H}-band best-fitting stars is by analysing their individual abundance ratios. For this, we need to derive the parameters of these stars using spectral template fitting with a synthetic spectral grid. However, at the very coolest temperatures, constructing stellar atmosphere models in the NIR is not a trivial task, because of the multiplicity of atomic lines and molecules. The MARCS \citep{gustafsson2008}  models provide a good representation of the atmospheres of late-type stars down to effective temperatures of about $3000$~{\small K}. These models have been chosen for  the APOGEE survey \citep{majewski2017}, and provide good agreement with the IR data, particularly for giant stars. We used the low-resolution spectra included with this collection of models to create a spectral library with a resolving power of 2000, matching that of the IRTF library.

We built two regular libraries, with equidistant steps in the atmospheric parameters. The first library spans
$2500  \le$ \Teff $\le 4000$~{\small K} in steps of $100$~{\small K}, and $-0.5 \le $\logg $\le 5.5$ in steps of $0.5$~dex. The second library spans $3500  \le$ \Teff $\le 6000$~{\small K} in steps of $250$~{\small K}, and $0.0 \le $\logg $\le 5.5$ in steps of $0.5$~dex. In addition to \Teff\ and \logg\ both libraries include two more parameters and \Feh, which spans $-2.5 \le $\Feh$ \le 1.0$ in steps of $0.5$~dex. Both libraries span the full wavelength range of the IRTF spectra ($8000-50000$~\AA).

We fit the spectra of the stars in Tab.~\ref{tab:tab1} using 
FERRE\footnote{\url{http://github.com/callendeprieto/ferre}}, deriving the optimal set of parameters that best represented the observations. We preserved the SED in  the observations and simply normalized the fluxes to have an average flux of $1$, and  fitted nearly the spectra between $8150$ and $27187$~\AA, where there is most information. An attempt to include longer wavelengths returned significantly poorer fits.

\begin{figure*}
\centering
	\includegraphics[width=\linewidth]{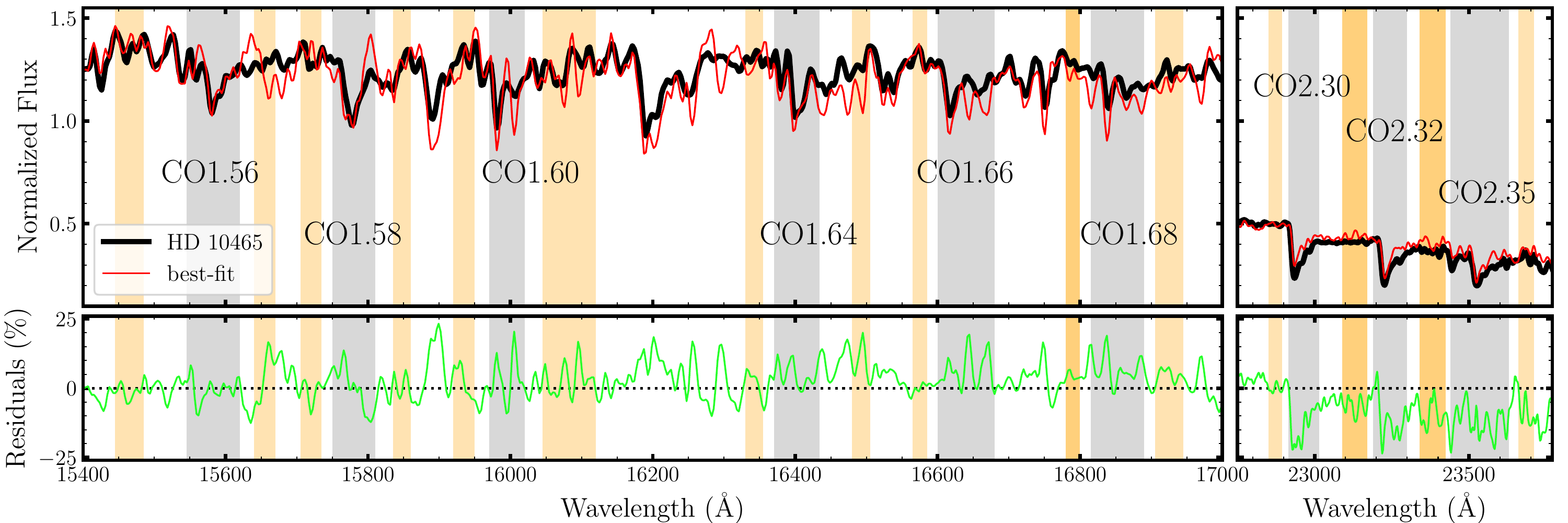}
    \caption{The upper panel shows the spectrum of one of the \textit{H}-band best-fitting stars, HD~10465, (black) in \textit{H} and \textit{K} bands, together with the best-fitted spectrum from MARCS library of very cool stars (red). CO features, with central absorptions and their pseudo-continua, are plotted as grey and orange shaded areas. The lower panel shows the relative residuals of the observed spectrum with respect to the best-fitted spectrum.}
    \label{fig:figB}
\end{figure*}

Figure~\ref{fig:figB} shows the best fitted spectrum (red line) to one of the stars (black line), in CO dominated regions, as an example. It is clear that the best fit is relatively poor, with residuals at a level of $\sim25$\% in the regions of CO absorption. The derived parameters of the stars are listed in the Tab.~\ref{tab:tabB}. The two stars in CO-strong sequence (HD~10465 and HD~187238) have lower $\alpha$ abundances than their counterparts (with the same \Teff) in the normal sequence (HD~36003, HD~219734); this is consistent with the behaviour of CO line-strengths in the $\alpha$-enhanced E-MILES SSP models that decrease with increasing the \aFe. However, the derived parameters of these stars with MARCS models are significantly different than the ones from the literature (Tab.~\ref{tab:tab1}), ($\Delta$\Teff\ $\sim-243$~{\small K}, $\Delta$\logg\ $\sim0.2$, $\Delta$\Feh\ $\sim0.1$). This experiment clearly shows the current limitations of stellar atmospheric models in the low-temperature regime.

\strutlongstacks{T}
\begin{table}
	\centering
	\caption{ Derived parameters of the \textit{H}-band best-fitting stars from MARCS grid for cool stars.  Columns~1 gives the name of each star. The derived $\alpha$ abundance, effective temperature, surface gravity, and  metallicity of the stars are given in Columns~2, 3, 4, and 5, respectively.  }
	\label{tab:tabB}
	\begin{tabular}{lcccr} 
		\hline\hline
		
		Star & \aFe & \Teff & \logg	& \Feh \\
		     & (dex) & ({\small K})	         & (dex)	& (dex) \\
		(1) & (2)	& (3)	& (4)	& (5) \\
	    \hline
	    
HD 219734 & 0.26  & 3720 & 0.15 & -0.35  \\

HD 36003 & 0.00 & 4578 & 5.02 & -0.57 \\

HD 187238 & -0.05 & 3604 & 1.77 & 0.77 \\

HD 10465 & -0.01 & 3590 & 0.50 & 0.60  \\
        
		\hline
	\end{tabular}
\end{table}


\bsp	
\label{lastpage}
\end{document}